\documentclass[preprint,showpacs,preprintnumbers,amsmath,amssymb]{revtex4}

\usepackage{graphicx}
\usepackage{dcolumn}
\usepackage{bm}

\def\slsh{\rlap{$\;\!\!\not$}}
\newcommand\pg         {p_g}
\newcommand\pb         {p_b}
\newcommand\pt         {p_t}
\newcommand\pw         {p_W}
\newcommand\tpW        {\tilde p_W}
\newcommand\tpb        {\tilde p_b}

\newcommand\nn         {\nonumber}

\newcommand\as         {\ensuremath{\alpha_{\mathrm{s}}}}

\newcommand\Oe[1]      {\ensuremath{\mathrm O(\ep^{#1})}}
\newcommand{\bV}       {{\bf V}}
\newcommand{\cV}       {{\cal V}}

\def\MW{M_W}
\def\ep{\epsilon}
\def\beq{\begin{equation}}
\def\eeq{\end{equation}}
\def\beqn{\begin{eqnarray}}
\def\eeqn{\end{eqnarray}}

\def\cM{{\cal M}}

\def\ket#1{|{#1}\rangle}
\def\bra#1{\langle{#1}|}

\def\bom#1{{\mbox{\boldmath $#1$}}}
\def\to{\rightarrow}

\def\nn{\nonumber}

\def\ID{1 \kern -.45 em 1}

\newcommand{\tpij}{\widetilde p_{ij}}
\newcommand{\tpk}{\widetilde p_k}
\newcommand{\zi}{\tilde z_i}
\newcommand{\zj}{\tilde z_j}

\newcommand{\yijk}{y_{ij,k}}
\newcommand{\vijk}{v_{ij,k}}

\newcommand{\tvijk}{\tilde v_{ij,k}}

\newcommand{\rd}{{\mathrm{d}}}

\newcommand{\eik}{{\mathrm{eik}}}
\newcommand{\coll}{{\mathrm{coll}}}

\newcommand{\CF}{C_{\mathrm{F}}}

\def\Li{\mathop{\mathrm{Li}}\nolimits}

\def\mathswitchr#1{\relax\ifmmode{\mathrm{#1}}\else$\mathrm{#1}$\fi}

\def\draftdate{\relax}
\def\mda{\relax}
\def\mua{\relax}
\def\mla{\relax}
\def\draft{
\def\thtystars{******************************}
\def\sixtystars{\thtystars\thtystars}
\typeout{}
\typeout{\sixtystars**}
\typeout{* Draft mode!
         For final version remove \protect\draft\space in source file *}
\typeout{\sixtystars**}
\typeout{}
\def\draftdate{August 13, 2004}
\def\mua{\marginpar[\boldmath\hfil$\uparrow$]%
                   {\boldmath$\uparrow$\hfil}%
                    \typeout{marginpar: $\uparrow$}\ignorespaces}
\def\mda{\marginpar[\boldmath\hfil$\downarrow$]%
                   {\boldmath$\downarrow$\hfil}%
                    \typeout{marginpar: $\downarrow$}\ignorespaces}
\def\mla{\marginpar[\boldmath\hfil$\rightarrow$]%
                   {\boldmath$\leftarrow $\hfil}%
                    \typeout{marginpar: $\leftrightarrow$}\ignorespaces}
\overfullrule 5pt
\oddsidemargin -15mm
\marginparwidth 29mm
}

\def\stars{\strut\leaders\hbox{*}\hfill\strut}
\def\starline{\hfil\strut\hfil\hbox to \textwidth {\stars}\hfil}

\def\bentarrow{\:\raisebox{1.3ex}{\rlap{$\vert$}}\!\rightarrow}
\def\dk#1#2#3{
\begin{array}{r c l}
#1 & \rightarrow & #2 \\
 & & \bentarrow #3
\end{array}
}

\def\bothdk#1#2#3#4#5{
\begin{array}{r c l}
#1 & \rightarrow & #2#3 \\
 & & \:\raisebox{1.3ex}{\rlap{$\vert$}}\raisebox{-0.5ex}{$\vert$}%
\phantom{#2}\!\bentarrow #4 \\
 & & \bentarrow #5
\end{array}
}
\begin{document}
\preprint{ANL-HEP-PR-04-70}
\preprint{FERMILAB-Pub-04/134-T}
\preprint{DSF-22/2004}
\preprint{hep-ph/0408158}
\title{Single top production and decay at next-to-leading order}
\author{John Campbell}
\email{johnmc@hep.anl.gov}
\affiliation{
High Energy Physics Division, Argonne National Laboratory,\\
Argonne, IL 60439, USA }
\altaffiliation[Address after October 1, 2004:]{ Department of Physics, 
 TH Division, CERN, CH-1211 Geneva 23, Switzerland}
\author{R. K. Ellis}%
\email{ellis@fnal.gov}
\affiliation{
Theoretical Physics Department, Fermi National Accelerator Laboratory,\\
P.~O.~Box 500, Batavia, IL 60510,  USA}

\author{Francesco Tramontano}
\email{Francesco.Tramontano@na.infn.it}
\affiliation{Dipartimento di Scienze Fisiche, Universit\`a di Napoli,\\
Complesso di Monte S. Angelo, Napoli, Italy}

\date{\today}

\begin{abstract}
We present the results of a next-to-leading order analysis of single
top production including the decay of the top quark.  Radiative
effects are included both in the production and decay stages, using a
general subtraction method. This calculation gives a good treatment of
the jet activity associated with single top production. We perform an
analysis of the single top search at the Tevatron, including a
consideration of the main backgrounds, many of which are also
calculated at next-to-leading order.
\end{abstract}

\maketitle

\section{Introduction}

Following the discovery of the top quark at the Tevatron in 
Run I~\cite{Abe:1994st,Abe:1995hr,Abachi:1995iq},
one of the aims of the current round of data-taking is to study the
top quark in more detail. In addition to the accumulation of further
statistics in the $t \bar{t}$ pair-production channel, both
collaborations are performing a search for single top 
production~\cite{Juste:2004an,Abbott:2000pa,Abazov:2001ns,Acosta:2004er,Acosta:2001un}. Since single top production proceeds by 
the exchange (or production) of a $W$ boson,
it offers another window into the weak interactions of the top quark and
potentially can lead to a direct measurement of $V_{tb}$.
Relative to the $t \bar{t}$ pair-production channel, single top production
is suppressed by the weak coupling but favored by phase space.

We shall report here on the method of inclusion of single top processes
into the general next-to-leading order Monte Carlo program 
MCFM~\cite{Campbell:1999ah,Campbell:2000bg,Campbell:2002tg}
and give phenomenological results relevant to the search strategy in Run II.
At the Born level the single top processes which we include are the 
$s$-channel process~\cite{Cortese:fw,Stelzer:1995mi,Heinson:1996zm},
\beq
u + {\bar d} \to W^* \to t + {\bar b}, 
\label{eq:schannel}
\eeq
the $t$-channel 
process~\cite{Willenbrock:cr,Yuan:1989tc,Ellis:1992yw,Carlson:1993dt,Heinson:1996zm},
\beq
b + u  \to t + d,
\label{eq:tchannel}
\eeq
and the $tW$ mode~\cite{Tait:1999cf,Belyaev:2000me},
\beq \label{eq:tWchannel}
b + g \to t + W^-.
\eeq
At the Tevatron, the $tW$ mode, Eq.~(\ref{eq:tWchannel}), evaluated 
in the Born approximation represents less than one percent of the total single 
top cross section; in this paper we will not consider
it further.  The processes shown in Eqs.~(\ref{eq:schannel}-\ref{eq:tWchannel})
are schematic. The actual implementation in the program includes 
a sum over all contributing partons in the initial state. In addition, we also 
include the leptonic decay of the top quark,
\beq
t \to \nu + e^+ + b,
\eeq
which allows a better comparison with experimental studies.

Although some information can be extracted from Born-level calculations,
the first serious approximation in QCD is obtained by including $O(\alpha_S)$
radiative corrections. We shall refer to this as next-to-leading order, (NLO).
It is only in NLO that a calculation gives any information about the choice
of factorization and renormalization scale. 
Moreover, if the calculation includes jets in the final state, it is only in
NLO that one obtains first information about the structure of the jets. 
The NLO corrections to the inclusive $s$-channel mode have been presented 
in ref.~\cite{Smith:1996ij} and the corrections to the inclusive 
$t$-channel process
have been considered in~ref.~\cite{Bordes:1994ki,Stelzer:1997ns}.
The NLO corrections to the differential distributions for the production
of single top, (without the decay of the top quark), 
which are needed for comparison with experimental results, 
have first been considered in ref.~\cite{Harris:2002md} for
the processes in Eqs.~(\ref{eq:schannel}) and~(\ref{eq:tchannel}).
For a recent update of this work see ref.~\cite{Sullivan:2004ie}.

In this paper we extend this program by
adding the leptonic decay of the top quark with full spin correlations,
as noted above, and also by
including the effects of gluon radiation in the decay.  
The approximations employed in incorporating the QCD corrections are
described in Section~\ref{qcdcorrections}.

Section~\ref{topdecay} outlines the calculation of the $O(\as)$ corrections 
to the decay of a free top quark which is helpful to establish notation.
The subtraction method which we use to 
include the radiative corrections to the decay of a top quark
in the single top production processes is described in 
Section~\ref{implementation}.

The search for single top is expected to be more challenging than
top-antitop associated production, both because of the smaller cross
section and the presence of larger backgrounds. We therefore give a
full account of the signal and background processes in Section~\ref{pheno}. 
Both the signal and the dominant background processes are evaluated at
next-to-leading order.

\section{QCD corrections}
\label{qcdcorrections}
\begin{figure}
\begin{center}
\includegraphics[angle=270,width=\columnwidth]{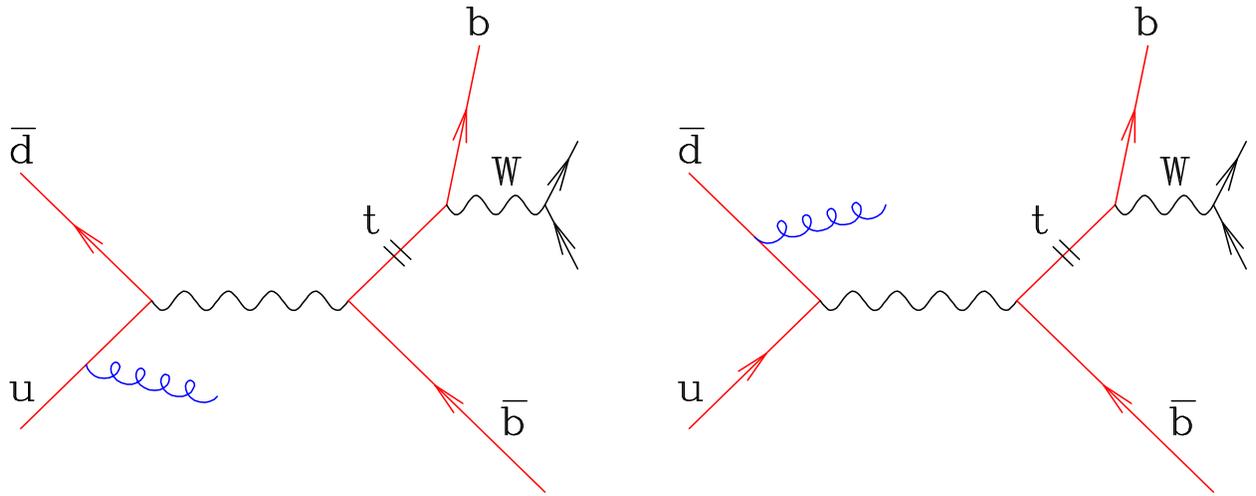}
\caption{Initial-state radiation in the production stage.
\label{fig:initialrad}}
\end{center}
\end{figure}

\begin{figure}
\begin{center}
\includegraphics[angle=270,width=\columnwidth]{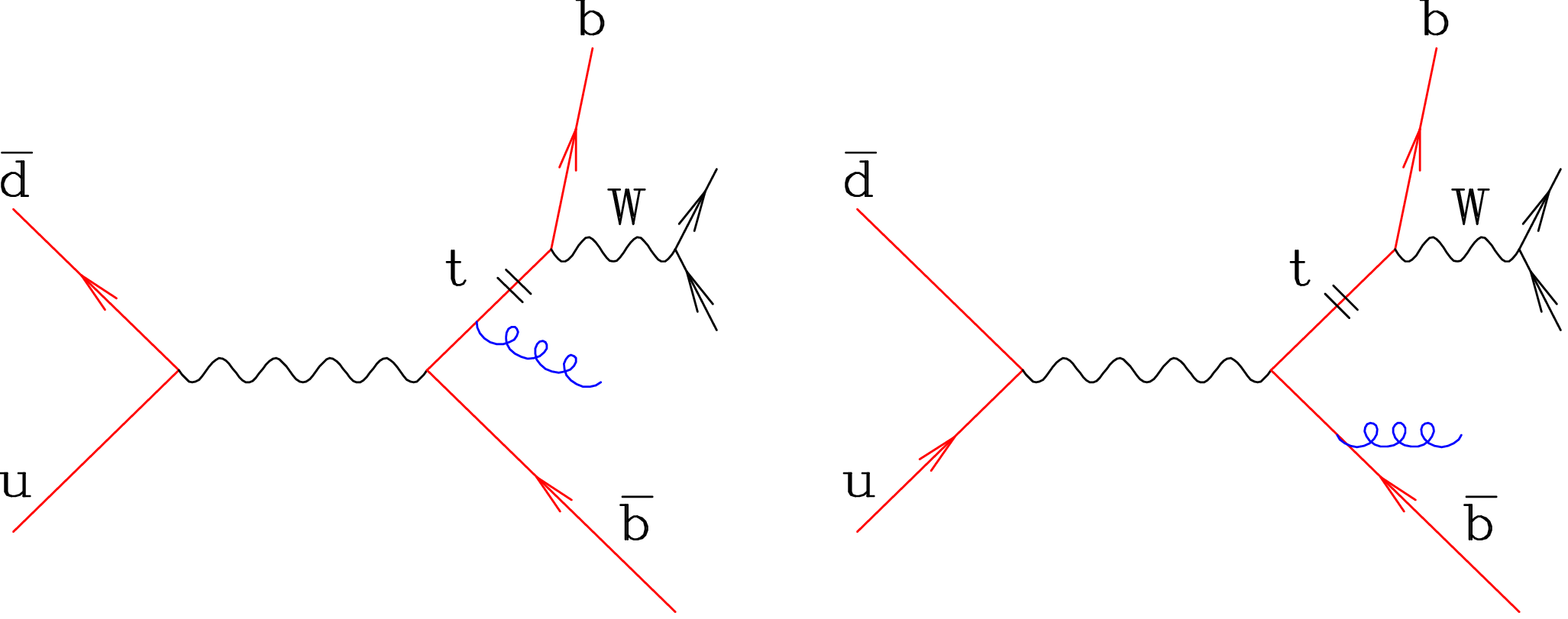}
\caption{Final-state radiation in the production stage.
\label{fig:prodrad}}
\end{center}
\end{figure}

\begin{figure}
\begin{center}
\includegraphics[angle=270,width=\columnwidth]{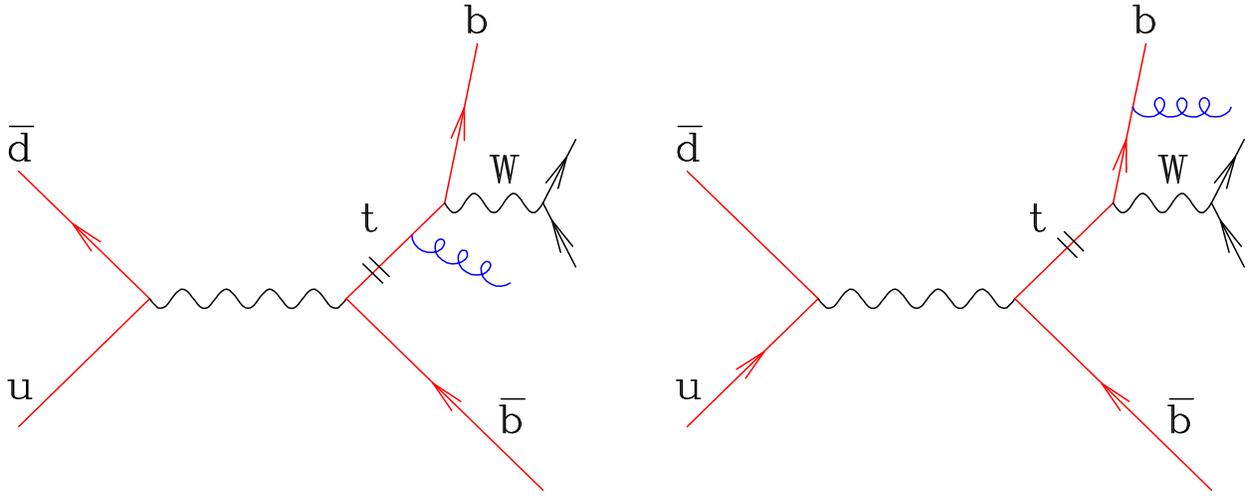}
\caption{Final-state radiation in the decay stage.
\label{fig:decayrad}}
\end{center}
\end{figure}

\begin{figure}
\begin{center}
\includegraphics[angle=270,scale=0.4]{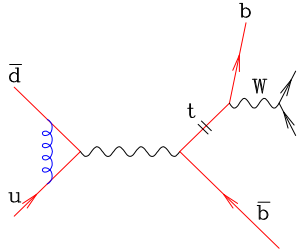}
\caption{Virtual radiation in the initial state.
\label{initv}}
\end{center}
\end{figure}

\begin{figure}
\begin{center}
\includegraphics[angle=270,width=\columnwidth]{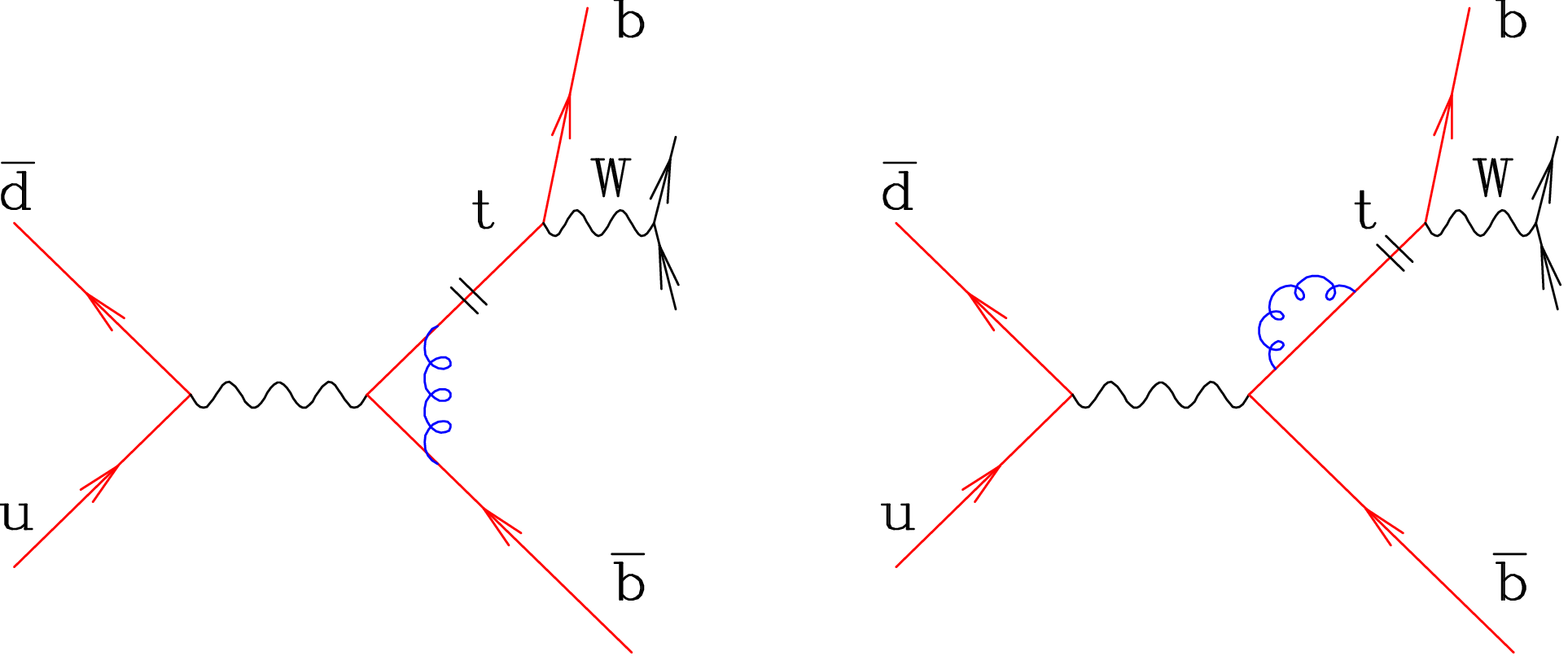}
\caption{Final-state virtual radiation in the production stage.
\label{fig:interv}}
\end{center}

\end{figure}

\begin{figure}
\begin{center}
\includegraphics[angle=270,width=\columnwidth]{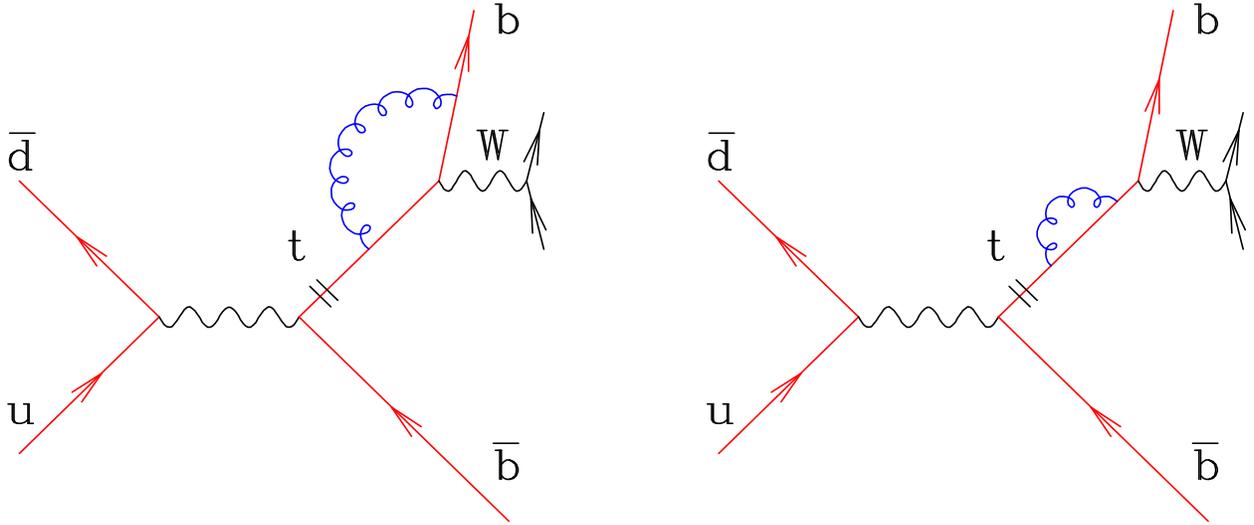}
\caption{Final-state virtual radiation in the decay stage.
\label{fig:finalv}}
\end{center}

\end{figure}

In order to describe the inclusion of radiative corrections we shall
discuss the $s$-channel process. The diagrams for the $t$-channel
process can easily be constructed by crossing. They are obtained by
reading the diagrams shown in 
Figs.~\ref{fig:initialrad}-\ref{fig:finalv} from bottom to top,
instead of from left to right. Some of the statements made below are
specific to the $s$-channel process, but the extensions to the $t$-channel
process are obvious.

We shall work in the on-shell approximation for the top quark. 
Thus every diagram considered has one top quark exactly on its 
mass shell. Diagrams without an on-shell top quark 
are suppressed by $\Gamma_t/m_t$
where $\Gamma_t$ and $m_t$ are the width and mass of the top quark.
In this approximation the real radiative corrections fall
into three types: radiation associated with the initial
state~(Fig.~\ref{fig:initialrad}), radiation in the final state
associated with the production of the top
quark~(Fig.~\ref{fig:prodrad}), and radiation in the final state
associated with the decay of the top
quark~(Fig.~\ref{fig:decayrad}). The double bars in these figures
indicate which top quark is on its mass shell. By producing the top
quark strictly on its mass shell we are assured that 
the diagrams in Figs.~\ref{fig:prodrad} and~\ref{fig:decayrad}
are separately gauge invariant. 


In addition to these real radiation  diagrams we also have virtual
radiation diagrams. In our approximation  these again fall into the
same three categories; virtual radiation  in the initial state
Fig.~{\ref{initv}}, final-state  virtual radiation in the production
stage Fig.~{\ref{fig:interv}}, and final-state  virtual radiation in
the decay stage, Fig.~{\ref{fig:finalv}}. Self energy contributions on
massless external lines have not been displayed. They give rise  to
scaleless integrals, which are set equal to zero in the dimensional
regularization scheme.


\begin{figure}
\begin{center}
\includegraphics[width=\columnwidth]{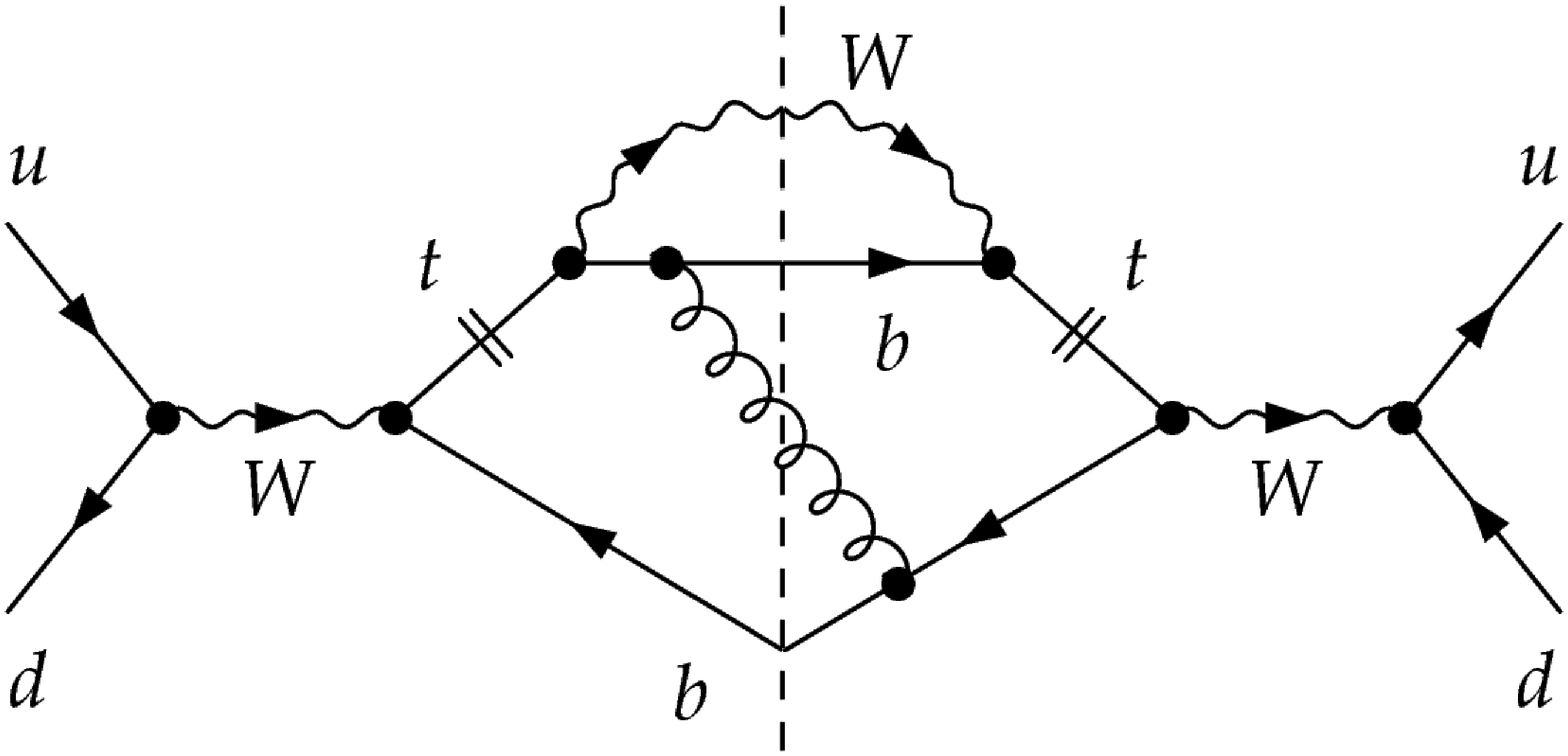}
\caption{Interference between radiation in production and decay, real terms.
\label{fig:REAL}}
\end{center}
\end{figure}

\begin{figure}
\begin{center}
\includegraphics[width=\columnwidth]{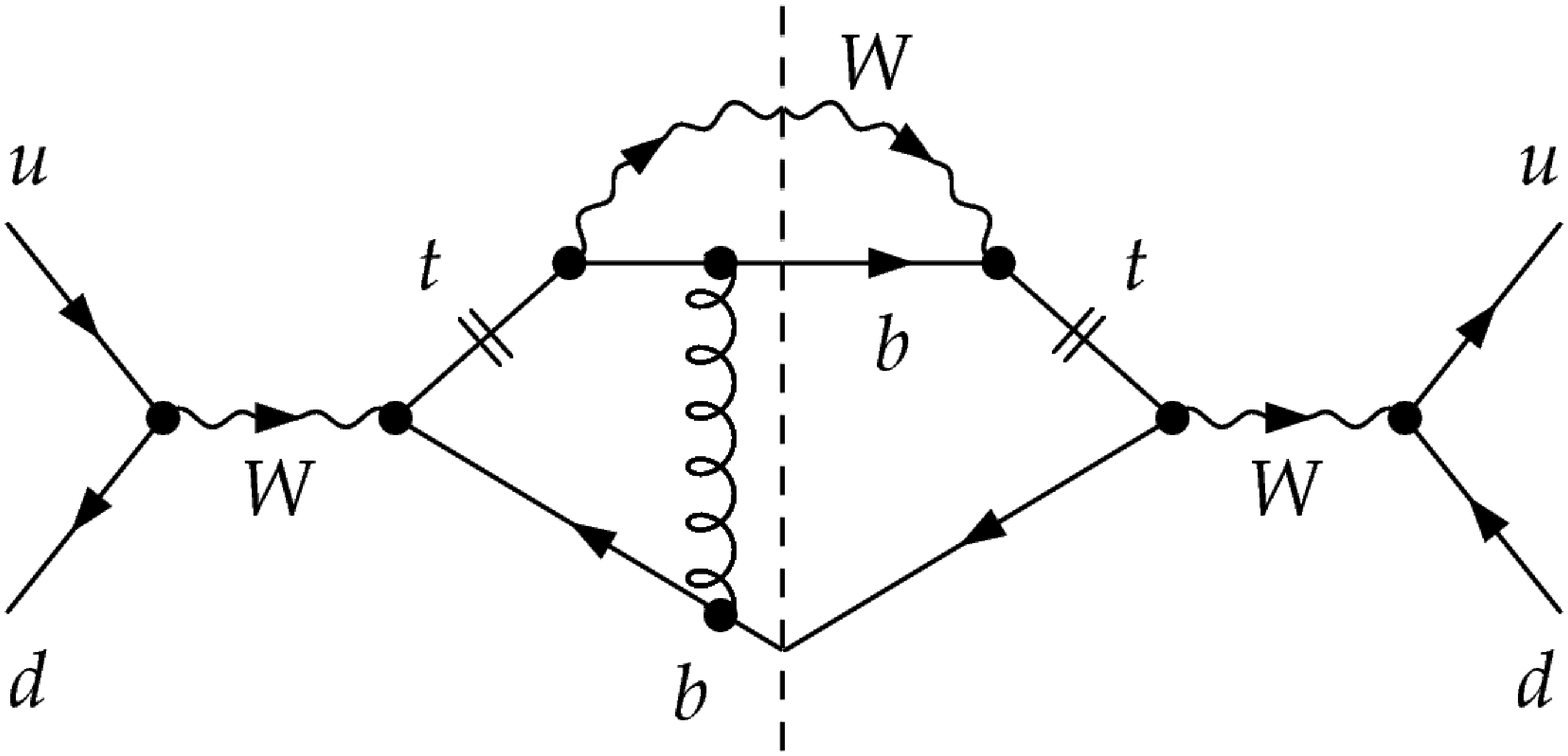}
\caption{Interference between radiation in production and decay, virtual terms
\label{fig:VIRT}}
\end{center}
\end{figure}

We have neglected the interference between real radiation in
production  and decay diagrams. An example is shown in
Fig.~\ref{fig:REAL}. We also neglect the virtual radiation  diagrams
that link the production and decay stages, Fig.~\ref{fig:VIRT}.
The physical reason for the neglect of these diagrams has been provided
in Refs.~\cite{Fadin:1993kt,Fadin:1993dz,Melnikov:np}. The
characteristic time scale for the production of the $t{\bar b}$ pair is
of order $1/m_t$ while the time for the decay is $1/\Gamma_t$.
Therefore in general, radiation in the production and decay stages are
separated by a large time and the interference effects average to zero.
The potentially dangerous region for this argument is the one in which
the emitted gluon is soft since this effect is not confined to a time
of order $1/m_t$. The phase space for soft radiation is limited
by the region in which the propagators remain resonant and the gluon
energy is less than of order $\Gamma_t$. However because of the
cancellation of real and virtual radiation the region of soft radiation
is not especially privileged.  Thus for infrared safe variables these
interference effects are expected to be of order $\as \Gamma_t/m_t$.

Confirmation of this suppression for the $s$-channel process
is provided by the work of Pittau~\cite{Pittau:1996rp}, who included
interference between the decay and production stages. The final results
are consistent with an effect of order $\Gamma/m_t$. A similar study
has been performed for $e^+e^- \to t {\bar t}$ including the
subsequent decay of the top quarks~\cite{Macesanu:2001bj}. Here 
the effect of nonfactorizable corrections in the invariant mass
distribution of the top was found to be very small.

The implementation of the cancellation of soft and collinear radiation
contributions which are separately divergent is performed using the
subtraction method~\cite{Ellis:1980wv}.  In our program MCFM, we have
consistently used the dipole subtraction method for massless particles
as developed by Catani and
Seymour~\cite{Catani:1996jh,Catani:1997vz}. For the case of single
top production we have a massive quark in the final state, so we have
implemented a generalization of this scheme as suggested
in~\cite{Catani:2002hc}. A useful further generalization has been
suggested by Nagy and Tr\'ocs\'anyi~\cite{Nagy:1998bb,Nagy:2003tz},
who introduced a tunable parameter $\alpha$ which controls the size of
the subtraction region. We have extended the massive results of
ref.~\cite{Catani:2002hc} to include this parameter.  Further details
may be found in Appendix~\ref{app:alphadep}.

In order to deal with radiation in the 
decay stage of the process we have developed a specialized 
subtraction procedure, which we discuss in Section~\ref{implementation}.
We expect that this method may be useful in other contexts.  In particular, 
it could be applied to the decay of the top quarks in 
the $t \bar{t}$ production process.

\section{Radiative corrections to top decay}
\label{topdecay}
One of the new results in this paper is the inclusion of the QCD corrections
to the decay of a top quark in a Monte Carlo program. Since the
phase space for the decay of an on-shell top quark factorizes from the
production phase space, most of the features of the full calculation
are present in the decay of an isolated top quark.  
For simplicity, and in order
to introduce the notation, we first reproduce the well-known result for
radiation from a free top quark~\cite{Jezabek:1988iv}. 

The lowest order process is $t(\pt) \rightarrow W(\pw)+b(\pb)$, 
where the momenta carried by the fields are shown in brackets.
The matrix element 
summed and averaged over initial spin and color is,
\beq
\overline{\sum} |{\cM}_{0}|^2 = 
 2 \frac{G_F}{\sqrt{2}} m_t^4 (1-r^2) (1+2r^2) \; .
\eeq
where we have defined $r^2=(\pt-\pb)^2/m_t^2$.
For the case of an on-shell $W$ boson, $r^2=\MW^2/m_t^2$. 
Here and throughout this paper, the mass of the $b$-quark is set equal to zero.
The corresponding Born-approximation width is given by,
\beq
\Gamma_{0} = \frac{G_F m_t^3}{8 \pi \sqrt{2} } (1-r^2)^2(1+2 r^2)
\label{eq:widthborn}
\eeq

\subsection{Virtual corrections}
\begin{figure}
\begin{center}
\includegraphics[angle=270,width=\columnwidth]{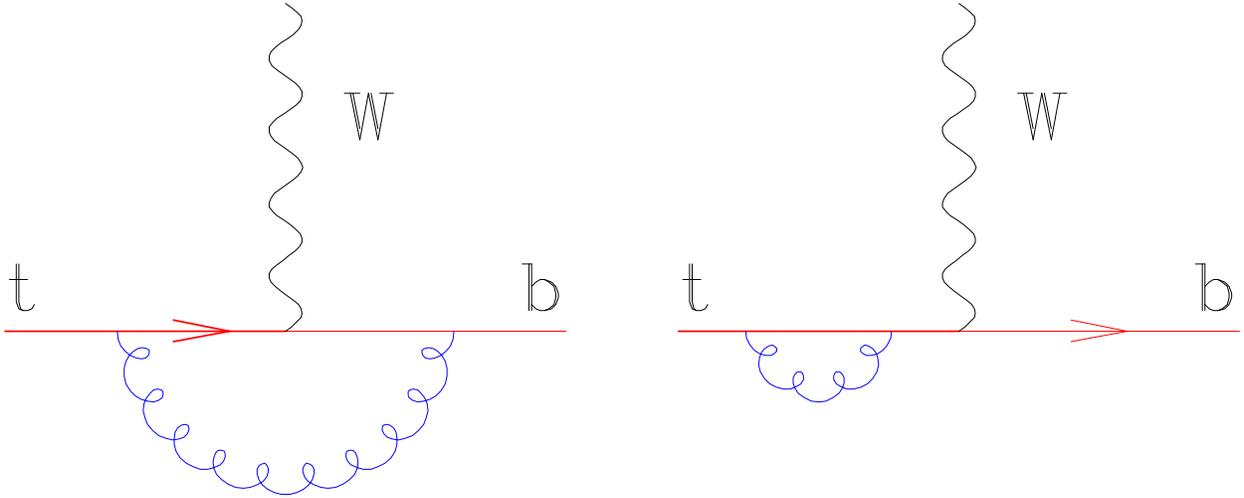}
\caption{Virtual radiation in free top decay}
\label{fig:freev}
\end{center}
\end{figure}

The form-factor for the process $t \to W + b$ including virtual gluon
corrections is defined by,
\beq
\Gamma^\mu(\pb,\pt)=\bar{u}(\pb) F^\mu(\pb,\pt) u(\pt)
\eeq
with a massless $b$ quark and a massive $t$ quark and momentum transfer 
$\pw=\pt-\pb$. The contributing diagrams are 
shown in Fig.~\ref{fig:freev}.
The result for the form factor evaluated through order $\as$ is 
well known~\cite{Gottschalk:1980rv,Schmidt:1995mr,Harris:2002md},
\beqn
F^\mu(\pb,\pt)&=&\gamma^\mu \gamma_L \Big[ 1 +
\frac{\alpha_S \; C_F }{4 \pi \Gamma(1-\ep)}  
\Big(\frac{4 \pi \mu^2}{m_t^2-i \varepsilon}\Big)^\ep 
C_0 \Big] \nn \\
&+&\frac{\alpha_S\; C_F}{4 \pi}\Big[
C_1 \frac{\pb^\mu}{m_t} \gamma_R 
+C_2 \frac{\pw^\mu}{m_t} \gamma_R \Big]
 +O(\ep)
\eeqn
where
\beqn
C_0&=&\Big\{-\frac{1}{\ep^2}
 -\frac{1}{\ep}\Big(\frac{5}{2}-2 \ln(1-r^2)\Big)
          - \frac{11+\eta}{2}-\frac{\pi^2}{6}
          - 2 \Li_2(r^2) \nn \\
   &+& 3 \ln(1-r^2) -2 \ln^2(1-r^2)- \frac{1}{r^2} \ln(1-r^2) \Big\} \nn \\
C_1&=& \frac{2}{r^2} \ln(1-r^2)
\label{eq:massivevert}
\eeqn
and 
\beq
\gamma_{R/L}= \frac{1}{2} (1 \pm \gamma_5) \; .
\eeq
The $C_2$ term will not contribute to physical amplitudes.
The ultra-violet and infra-red divergences have been 
regulated by continuing to $d=4-2 \ep$ dimensions.
The dimensional regularization scheme is determined by the parameter
$\eta$ as follows,
\beqn
\label{eq:etadef}
\eta &=& 1,\;\;\mbox{'t Hooft-Veltman scheme~\cite{'tHooft:1972fi}} \nn \\ 
\eta &=& 0,\;\;\mbox{four-dimensional helicity scheme~\cite{Bern:2002zk}}\;.
\eeqn
The final result for the 
virtual correction to the total decay width is,
\beq \label{eq:virtbit}
\Gamma_{\rm virtual} = \Gamma_{0} \;
 \frac{\alpha_S C_F}{2\pi \Gamma(1-\ep)}
 \left( \frac{4\pi \mu^2 }{m_t^2} \right)^{\ep}  
 \left( C_0 + \frac{1}{2} C_1 \frac{1-r^2}{1+2r^2} \right) +O(\ep) \; .
\eeq

We now make three parenthetic remarks which are useful for including 
the virtual corrections in the production processes, 
Eqs.~(\ref{eq:schannel},\ref{eq:tchannel}).
First, we note that Eq.~(\ref{eq:massivevert}) 
is valid for $r^2 < 1$. For the case when $r^2 > 1$, 
which is needed for the virtual
corrections to the $s$-channel production process,
we apply the transformation,
\beq
\Li_2(r^2)=\frac{\pi^2}{6}-\Li_2(1-r^2) -\ln(1-r^2) \ln r^2 
\eeq
and analytically continue by replacing $\ln(1-r^2)$ with 
$\ln(1-r^2-i\varepsilon)$.
Second, we note that the result for the
light quark vertex including the virtual corrections, which is also needed
for the case of production, is as follows~\cite{Altarelli:1979ub},
\beq
\Gamma^\mu(p_d, p_u)=\bar{u}(p_d) \gamma^\mu \gamma_L u(p_u) \; 
\Bigg[ 1 + \frac{\alpha_S C_F }{4 \pi \Gamma(1-\ep)}  
\Big(\frac{4 \pi \mu^2}{-q^2-i \varepsilon}\Big)^\ep
\Big\{-\frac{2}{\ep^2}-\frac{3}{\ep}-7-\eta+O(\ep) \Big\}\Bigg]
\eeq
with $q=p_u-p_d$. Third, for the case of the production processes
the singularities in these virtual corrections are canceled by equal
and opposite singularities in the integrated dipoles.
With a little work, using results for the integrated dipole terms in 
Appendix~\ref{app:alphadep}, one can demonstrate this cancellation.

\subsection{Real corrections}

\begin{figure}
\begin{center}
\includegraphics[angle=270,width=\columnwidth]{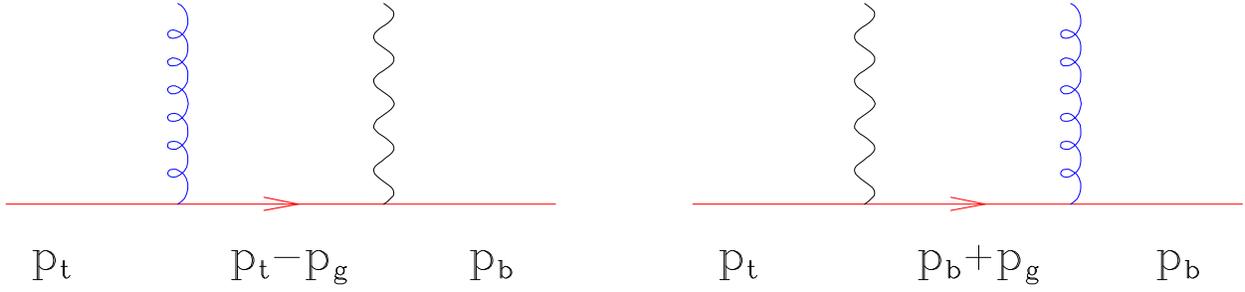}
\caption{Real radiation in free top decay}
\label{fig:free}
\end{center}
\end{figure}
The Feynman diagrams for the process
\beq
t(\pt) \rightarrow b(\pb)+W(\pw)+g(\pg) \; ,
\eeq
are shown in Fig.~\ref{fig:free}.
The result for the matrix element squared in four dimensions 
is given as follows
\beqn
\overline{\sum} |{\cM}|^2 &=& 
2 g^2 C_F  \frac{G_F}{\sqrt{2}} m_t^2
   \Bigg[m_t^2 (1-r^2) (1+2 r^2) 
\Big(2 \frac{\pt.\pb}{\pb.\pg \; \pt.\pg}-\frac{m_t^2}{(\pt.\pg)^2}\Big) \nn \\
       &+&8 r^2
      +2 \frac{(\pt.\pg-\pb.\pg)^2}{\pb.\pg \; \pt.\pg}(1+2 r^2)\Bigg] \; .
\eeqn
Applying the substitutions 
\beqn \label{eq:yztransformations}
\pb \cdot \pg &=& \frac{m_t^2}{2}(1-r)^2 y \nn \\
\pt \cdot \pg &=& \frac{m_t^2}{2} (1-r^2) (1-z) \; ,
\eeqn
the four-dimensional matrix element becomes,
\beqn
\label{eq:realmatrixelement}
\overline{\sum} 
|{\cM}|^2&=& 
  g^2 C_F \Bigg\{ \Big[ \frac{1}{\pb.\pg} 
 \Big(\frac{2}{1-z}-1-z\Big)-\frac{m_t^2}{(\pt.\pg)^2}\Big] \;
 \overline{\sum} |{\cM}_{0}|^2  \nn \\
  &+& 8 m_t^2 \frac{G_F}{\sqrt{2}} \Big[ 
 \frac{y}{1-z} \Big(\frac{3}{(1+r)}-r^2+2  r -\frac{5}{2}\Big)
  +\frac{(1+2 r^2)}{(1-z)}-1 \Big] \Bigg\}
\eeqn

In the rest frame of the top we obtain the two particle phase space 
for the decay $t \rightarrow W+ b$
\beqn
d \Phi^{(2)}(\pw,\pb;\pt) &=&  \frac{d^n \pb}{(2 \pi)^{n-1}}
\frac{d^n \pw}{(2 \pi)^{n-1}} \; (2 \pi)^n \; \delta^n(\pt-\pw-\pb) \; 
\delta(\pb^2) \; 
\delta(\pw^2-r^2 m_t^2) \nn \\
&=&=\frac{(4 \pi)^{2 \ep}}{8 \; (2 \pi)^2} 
\frac{1}{(m_t^{2})^\ep} \;
(1-r^2)^{1-2 \ep} d ^{n-1}\Omega_w
\label{eq:PS2}
\eeqn
The corresponding result in the rest frame of the top 
for the decay $t \rightarrow W + b + g$ is
\beqn
d \Phi^{(3)}(\pw,\pb,\pg;\pt)&=&
\frac{(1-r^2) (1-r)^{2-4\ep}}{32 \; (2\pi)^4} \; 
\; \frac{(m_t^2)^{1-2\ep} (4 \pi)^{3\ep}}{ 
\Gamma(1-\ep)}  \; d ^{n-1}\Omega_w \nn \\
&\times & \int_0^1 \; dz \;  
(r^2+z (1-r^2))^{-\ep} \; \int_0^{y_{max}} dy \;  
y^{-\ep} \; (y_{max} -y)^{-\ep} 
\label{eq:PS3}
\eeqn
where the upper integration limit $y_{max}$ is given by
\beq
y_{max}=\frac{(1+r)^2 z (1-z)}{\Big( z+r^2 (1-z)\Big)}.
\eeq

Taking the ratio of Eqs.~(\ref{eq:PS3}) and~(\ref{eq:PS2}) we
obtain the factorized form,
\beqn \label{eq:PSfactorized}
d \Phi^{(3)}(\pw,\pb,\pg;\pt)&=& d \Phi^{(2)}(\pw,\pb;\pt)
\times 
\frac{(1-r)^{2}}{16 \pi^2} \; 
\; \frac{(m_t^2)^{1-\ep} (4 \pi)^{\ep}}{
\Gamma(1-\ep)} \nn \\
&\times & \Big(\frac{1+r}{1-r}\Big)^{2 \ep} \; \int_0^1 \; dz \;  
(r^2+z (1-r^2))^{-\ep} \; \int_0^{y_{max}} dy \;  
y^{-\ep} \; (y_{max} -y)^{-\ep} \nn \\
\eeqn
Values of the reduced integrals defined by
\beq \label{integdef}
\Big\langle f(y,z) \Big\rangle = \Big(\frac{1+r}{1-r}\Big)^{2 \ep} 
\int_0^1 \; dz \;  
(r^2+z (1-r^2))^{-\ep} \; \int_0^{y_{max}} dy \;  
y^{-\ep} \; (y_{max} -y)^{-\ep} \; f(z,y)
\eeq
are given in Table~\ref{integtable}.
\renewcommand{\baselinestretch}{2.5}
\begin{table*}
\begin{centering}
\caption{Table of integrals defined by Eq.~(\ref{integdef}).\label{integtable}}
\begin{ruledtabular}
\begin{tabular}{ll}
$\Big\langle \frac{1}{y} \frac{2}{(1-z)}\Big\rangle $ 
& $\frac{1}{\ep^2}-\frac{2}{\ep} \ln(1-r^2)
+2 \ln^2(1-r^2) +2 \Li_2(1-r^2)-\frac{5 \pi^2}{6} + O(\ep)$\\
\hline
$\Big\langle \frac{1}{y} \Big\rangle 
$ & $-\frac{1}{\ep}-3 +2 \ln(1-r^2)
+\frac{r^2}{(1-r^2)} \ln r^2 + O(\ep)$  \\
\hline
$\Big\langle \frac{z}{y} \Big\rangle 
$ & $-\frac{1}{2 \ep}-\frac{(7-5 r^2)}{4 (1-r^2)}
+\ln(1-r^2) -\frac{r^4 }{2 (1-r^2)^2} \ln r^2  + O(\ep)$ \\
\hline
$\Big\langle \frac{-2}{(1+r)^2}\frac{1}{(1-z)^2}\Big\rangle 
$ & $\frac{1}{\ep}+2 -\frac{2 r^2}{1-r^2} \ln r^2 
 -2 \ln (1-r^2) + O(\ep)$ \\
\hline
$\Big\langle \frac{1}{(1+r)^2}\frac{1}{(1-z)}\Big\rangle 
$ & $\frac{r^2 \ln r^2}{(1-r^2)^2}+\frac{1}{1-r^2} + O(\ep)$ \\
\hline
$\Big\langle \frac{1}{(1+r)^4}\frac{y}{(1-z)}\Big\rangle 
$ & $\frac{r^2(2+r^2)}{2 (1-r^2)^4}\ln r^2
+\frac{(1+5 r^2)}{4 (1-r^2)^3} + O(\ep)$ \\
\hline
$\Big\langle (1-r)^2 \Big\rangle 
$ & $ \frac{(r^2+1)}{2}+\frac{r^2}{1-r^2} \ln r^2  + O(\ep)$\\
\end{tabular}
\end{ruledtabular}
\end{centering}
\end{table*}
\renewcommand{\baselinestretch}{1.0}

Using these integrals we can calculate the contribution
to the total decay width from the real diagrams. The result is,
\beqn \label{eq:realbit}
\Gamma_{\rm real} &=& \Gamma_{0}^{(d)} \;
 \frac{\alpha_S C_F}{2\pi \Gamma(1-\ep)}
 \left( \frac{4\pi \mu^2 }{m_t^2} \right)^{\ep}  
 \Bigg( \frac{1}{\ep^2} + \frac{1}{\ep}
  \left( \frac{5}{2} - 2\ln(1-r^2) \right) - \frac{5\pi^2}{6}
  +2\Li_2(1-r^2) \nn \\
  &&-5\ln(1-r^2)+2\ln^2(1-r^2)
  -\frac{2r^2(1+r^2)(1-2r^2)}{(1-r^2)^2(1+2r^2)} \ln r^2
  -\frac{2(7r^4-5r^2-4)}{(1+2r^2)(1-r^2)} + \frac{\eta}{2} \Bigg) \nn \\ \; .
\eeqn
The $\eta$-dependence has been restored in this equation, so that 
Eq.~(\ref{eq:realbit}) is valid also for the 't Hooft Veltman scheme.

\subsection{Result for the corrected width}

Using the virtual corrections of 
Eq.~(\ref{eq:virtbit}) 
and the real corrections of Eq.~(\ref{eq:realbit}) 
we can calculate the value of
the top width at $O(\alpha_S)$.
We write the correction to the decay rate in the form
\begin{equation}
\Gamma = \Gamma _{0}+\as \Gamma_{1}
\label{eq:correcteddecayrate}
\end{equation}
Our result is in agreement with the $m^2_b /m^2_t \to 0$ limit of the 
original calculation in ref.~\cite{Jezabek:1988iv},
\begin{eqnarray}
\frac{\as \Gamma_1}{\Gamma_0} &=& -\frac{\as}{2 \pi} C_F\Bigg[\frac{2}{3}\pi^2 +4 \;\Li_2(r^2)-\frac{3}{2}
 -2 \; \ln\Big(\frac{r^2}{1-r^2}\Big) +2 \; \ln r^2 \ln(1-r^2) 
\nn \\
&-&\frac{4}{3(1-r^2)}+\frac{(22-34 \; r^2)}{9 (1-r^2)^2}\ln r^2
       +\frac{(3+27 \ln(1-r^2)-4 \ln r^2 )}{9 (1+2 r^2)} \Bigg] \; ,
\end{eqnarray}
where $r=\MW /m_t$.
For $r \sim 4/9$ the QCD correction amounts to
\begin{equation}
\frac{\as \Gamma_{1}}{\Gamma_{0}} \approx -0.8 \as
\label{eq:correction_to_width}
\end{equation}
which lowers the leading order result for the width by about 10\%.  

\section{Factorization of singularities in top quark decay}
\label{implementation} 
We wish to construct a counter-term for the process
\beq
t \rightarrow W + b +g
\eeq
which has the same soft and collinear singularities as the 
full matrix element. This counter-term
takes the form of a lowest order matrix element multiplied by a
function $D$ which describes the emission of soft or collinear
radiation, 
\beq \label{decayfactorization}
|{\cM}( \ldots p_t,p_W,p_b,p_g) |^2 \to
|{\cM}_0( \ldots p_t,\tpW,\tpb) |^2 \times 
D(\pt.\pg,\pb.\pg,m_t^2,\MW^2) \; ,
\eeq
In the region of soft emission, or in the region where the momenta
$p_g$ and $p_b$ are collinear, the right hand side of 
Eq.~(\ref{decayfactorization}) 
has the same singularity structure as the full matrix element. 
The lowest order matrix element $\cM_0$ in Eq.~(\ref{decayfactorization}) 
is evaluated for values of the momenta $\pw$ and $\pb$
modified to absorb the four-momentum carried away by the gluon,
and subject to
the momentum conservation constraint, $\pt \rightarrow {\tpW}+\tpb$. 
The modified momenta denoted by a tilde are also subject to 
the mass-shell constraints, ${\tpb}^2=0$ and ${\tpW}^2=\pw^2$. 
The latter condition is necessary in order that the rapidly varying 
Breit-Wigner function for the $W$ is evaluated at the same kinematic point 
in the counterterm and in the full matrix element.
We define $\tpW$ by a 
Lorentz transformation, ${\tpW}^\mu=\Lambda^{\mu}_{\nu} \pw^\nu$
fixed in terms of the momenta $\pw$ and $\pt$.
Because $\tpW$ and $\pw$ are related by a Lorentz transformation 
the phase space for the subsequent
decay of the $W$ is unchanged.

The general form of a Lorentz transformation in the plane of
the vectors $\pt$ and $\pw$ is given by
(${\tpW}^\mu=\Lambda^{\mu}_{\nu} \pw^\nu$),
\beqn
\Lambda^{\mu\nu}&=&g_{\mu \nu}
     +\frac{\sinh(x)}{\sqrt{(\pt \cdot \pw)^2 -\pw^2 \pt^2}} \; 
\Big(\pt^{\mu} \pw^{\nu}-\pw^{\mu} \pt^{\nu} \Big) \nn  \\
 &+&\frac{\cosh(x)-1}{(\pt \cdot \pw)^2 -\pw^2 \pt^2} \;
      \Big(\pt \cdot \pw \; (\pt^\mu \pw^\nu+\pw^{\mu} \pt^{\nu})
    -\pw^2 \; \pt^{\mu} \pt^{\nu}-\pt^2 \; \pw^{\mu} \pw^{\nu}\Big)
\label{eq:lorentztrans}
\eeqn
The transformed momentum of the $b$ quark is fixed by $\tpb=p_t-\tpW$.
For the special case in which we impose the condition 
$\tpb^2=(\pt - {\tpW})^2=0$ we get
\beqn
\label{eq:lorentztransa}
\sinh(x)&=&\frac{1}{2 \; \pt^2 \pw^2} \Big[-(\pt^2-\pw^2) \pt\cdot \pw
 +(\pt^2+\pw^2) \sqrt{(\pt \cdot \pw)^2 -\pw^2 \pt^2}\; \Big]
 \nonumber \\
\cosh(x)&=&\frac{1}{2 \; \pt^2 \pw^2} \Big[+(\pt^2+\pw^2) \pt\cdot \pw
 -(\pt^2-\pw^2)\sqrt{(\pt \cdot \pw)^2 -\pw^2 \pt^2}\; \Big]
\eeqn
Acting on the vector $\pw$ the Lorentz transformation becomes
\beq  \label{eq:actingonpw}
\tpW = \alpha \; \big(\pw -\frac{\pt \cdot \pw}{\pt^2} \pt\big) +
\beta \; \pt 
\eeq
where the constants are given by
\beqn
\alpha & = & \frac{\pt^2-\pw^2}{2 \sqrt{(\pt \cdot \pw)^2-\pw^2 \pt^2}} \\
\beta & = & \frac{\pt^2+\pw^2}{2 \pt^2}
\eeqn
Eq.~(\ref{eq:actingonpw}) makes it clear that,
in the top rest frame, the transformation
on $\pw$ is a Lorentz boost along the direction 
of the $W$.

The momenta of the decay products of the $W$ can similarly 
be obtained by the same Lorentz transformation, 
Eqs.~(\ref{eq:lorentztrans},\ref{eq:lorentztransa}).

\subsection{Subtraction counterterm}
From Eq.~(\ref{eq:PSfactorized}) we may write the phase space
for the decay of an on-shell top quark as 
\beqn \label{eq:phasespacefact}
d \Phi^{(3)} (\pw,\pb,\pg;\pt) 
&=& d \Phi^{(2)} (\pw,\pb;\pt) 
 \int [dg(\pt,\pw,y,z)] \nn \\
&\equiv & d \Phi^{(2)} ({\tpW},{\tpb};\pt) 
 \int [dg(\pt,{\tpW},y,z)]
\eeqn
The equivalence in Eq.~(\ref{eq:phasespacefact}) 
follows from Eq.~(\ref{eq:PS2}) because
$d ^{n-1}\Omega_w= d ^{n-1}\Omega_{\tilde w}$ since $\pw$ and ${\tpW}$
are related by a boost. From Eq.~(\ref{eq:PSfactorized}) we see that the
phase space integral for the emitted gluon is given by,
\beqn
[dg(\pt,\tpW,y,z)] &=&
\frac{(1-r)^{2}}{16 \pi^2} \; 
\; (m_t^2)^{(1-\ep)} \; 
\frac{(4 \pi)^{\ep}}  {\Gamma(1-\ep)} \nn \\
&\times & \Big(\frac{1+r}{1-r}\Big)^{2 \ep} \; \int_0^1 \; dz \;  
(r^2+z (1-r^2))^{-\ep} \; \int_0^{y_{max}} dy \;  
y^{-\ep} \; (y_{max} -y)^{-\ep} 
\eeqn
where $y$ and $z$ are given by Eq.~(\ref{eq:yztransformations}).
By extension of Eq.~(\ref{eq:realmatrixelement}) 
we choose the counterterm to be,
\beq
D((p_t+p_g)^2,(p_b+p_g)^2,m_t^2,\MW^2) = g^2 \mu^{2 \ep} 
C_F \Bigg[ \frac{1}{\pb.\pg}\Big(\frac{2}{(1-z)}-1-z-\eta \ep (1-z)\Big) 
-\frac{m_t^2}{(t.g)^2}  
\Bigg]
\eeq
where the role of the parameter $\eta$ is defined in Eq.~(\ref{eq:etadef}). 
Performing the integral using the results of Table~\ref{integtable}, we obtain
the following result for the integrated counterterm,
\beqn
\int [dg(\pt,\tpW,y,z)] &\times& D((p_t+p_g)^2,(p_b+p_g)^2,m_t^2,\MW^2) 
= \nn \\
&&\frac{\alpha_S C_F}{2 \pi} 
\; \frac{(4 \pi \mu^2)^{\ep}}{m_t^{2 \ep} \Gamma(1-\ep)}
\Big[
      \frac{1}{\ep^2}
       + \frac{1}{\ep} \Big(\frac{5}{2} - 2 \ln(1-r^2) \Big) \nn \\
          &+& \frac{25}{4}
          + \frac{1}{2} \Big(\frac{1}{(1-r^2)^2}
          - \frac{8}{(1-r^2)}+7 \Big) \ln r^2
          + \frac{1}{2 (1-r^2)} \nn \\
          &+& 2 \Li_2(1-r^2)
          - \frac{5 \pi^2}{6}
          - 5 \ln(1-r^2)
          + 2 \ln^2(1-r^2)
          + \frac{\eta}{2}
\Big]
\eeqn

\section{Phenomenological studies}
\label{pheno}
The method of the previous section has been implemented in the 
Monte Carlo program, MCFM, allowing us to make predictions
for kinematic distributions of both signal and background events.
Before proceeding to describe our results for jets and the decay
products of the top, we present a number of results on total cross
sections.
\subsection{Total cross section}
Tables~\ref{tab:cteq6tot} 
and~\ref{tab:mrs02tot} give total cross sections using recent 
parton distributions and the updated top quark mass~\cite{Azzi:2004rc}. 
The input parameters which we use throughout this phenomenological section 
are presented in Table~\ref{tab:parameters}.
\begin{table}[tb]
\caption{\small LO and NLO cross sections for single top-quark
production at the Tevatron and LHC for $m_t=178$ GeV. 
The branching ratio for the decay of the top quark is not included. 
Cross sections are evaluated with CTEQ6L1 ($\as(M_Z)=0.130$) and CTEQ6M
($\as(M_Z)=0.118$) PDFs~\cite{Pumplin:2002vw}, 
and all scales set to $m_t$.  The errors represent Monte Carlo statistics only.
\label{tab:cteq6tot}}
\begin{center}
\begin{ruledtabular}
\begin{tabular}{llll} 
Process      & $\sqrt{s}$~[TeV] & $\sigma_{LO}$ [pb] & $\sigma_{NLO}$ [pb] \\
\hline
$s$-channel, $p\bar p$ ($t$) &1.96   & 0.270 & 0.405   $\pm$0.0003 \\
$s$-channel, $pp$ ($t$)      &14    & 4.26  & 6.06    $\pm$0.004 \\
$s$-channel, $pp$ ($\bar t $)&14    & 2.58  & 3.76    $\pm$0.003 \\
\hline
$t$-channel, $p\bar p$ ($t$) &1.96   & 0.826  & 0.924  $\pm$0.001 \\
$t$-channel, $pp$ ($t$)      &14    & 146.2  & 150.0  $\pm$0.2 \\
$t$-channel, $pp$ ($\bar t $)&14    & 84.8   & 88.5   $\pm$0.1 \\
\end{tabular}
\end{ruledtabular}
\end{center}
\end{table}
\begin{table}[tb]
\caption{\small LO and NLO cross sections for single top-quark
production at the Tevatron and LHC for $m_t=178$ GeV.  
The branching ratio for the decay of the top quark is not included. 
Cross sections
are evaluated with the MRST2002 NLO PDF set~\cite{Martin:2002aw}
with $\as(M_Z)=0.1197$, and all scales set to $m_t$.  
The errors represent Monte Carlo statistics only.
\label{tab:mrs02tot}}
\begin{center}
\begin{ruledtabular}
\begin{tabular}{llll} 
Process     & $\sqrt{s}$~[TeV] & $\sigma_{LO}$ [pb] &
 $\sigma_{NLO}$ [pb] \\
\hline
$s$-channel, $p\bar p$ ($t$) &1.96 & 0.285 & 0.404 $\pm$0.0003 \\
$s$-channel, $pp$ ($t$)       &14   & 4.57  & 6.17 $\pm$0.004 \\
$s$-channel, $pp$ ($\bar t $) &14   & 2.85  & 3.86 $\pm$0.003 \\
\hline
$t$-channel, $p\bar p$ ($t$)     &1.96   & 1.009 &  1.032    $\pm$0.001 \\
$t$-channel, $pp$ ($t$)          &14     & 160.1   & 154.4   $\pm$0.2 \\
$t$-channel, $pp$ ($\bar t $)    &14     & 96.9   & 92.4     $\pm$0.1 \\
\end{tabular}
\end{ruledtabular}
\end{center}
\end{table}
\begin{table}
\caption{Input parameters\label{tab:parameters}}
\begin{center}
\begin{ruledtabular}
\begin{tabular}{ll} 
$m_t=178.$~GeV  & $\Gamma_0=1.651$~GeV \\
$M_W=80.4$~ GeV  & $\Gamma_W =2.06$~GeV \\
$M_Z=91.188$~GeV  & $\Gamma_Z=2.49$~GeV \\
$G_F=1.16639\times 10^{-5}$~GeV$^{-2}$&  $\alpha^{-1}=132.351$ \\
$g_W^2=0.42651$       &   $e^2=0.0949475$
\end{tabular}
\end{ruledtabular}
\end{center}
\end{table}
We have not performed a full
analysis of the theoretical errors on these predictions.
It is important to remember that the results for the $t$-channel 
process depend on the $b$-quark parton distribution, 
which is calculated rather than measured. This is the source of the relatively
large difference between the $t$-channel Tevatron cross sections in
Tables~\ref{tab:cteq6tot} and~\ref{tab:mrs02tot}, compared to the
$s$-channel process which is dominated by well-constrained valence quark
distributions.

We have also compared our leading order (LO) and next-to-leading order
(NLO) results for the total cross sections 
(without radiation in the decay) at $\sqrt{s}=1.96$ and 14~TeV
with the results in ref.~\cite{Harris:2002md}. 
With the appropriate modification of the input parameters,
we find agreement with their results within the quoted errors. 

We now investigate what effect the inclusion of QCD corrections
in the decay of the top quark has on the total cross section.
Radiation in the decay should not change the total cross section.
However in a perturbative approach this means that the  difference
should be of higher order in $\as$.
Including QCD radiation effects only in the production, we obtain
\begin{equation}
\sigma B_{t \to b \nu e} \equiv \sigma B_{W \to \nu e}=
\sigma_{(0)}B_{W \to \nu e}+ \as \sigma_{(1)} B_{W \to \nu e}
\end{equation}
where $\sigma_{(0)}$ and $\sigma_{(1)}$ are the LO and NLO cross sections,
because in this order the width $t \to bW$ is equal to the total width.
When we also  include radiative corrections to the decay, we use the 
radiatively corrected total width $\Gamma=\Gamma_0+\as \Gamma_1$,
cf. Eqs.~(\ref{eq:widthborn},\ref{eq:correcteddecayrate}),
\begin{equation}
\sigma B_{t \to b \nu e+X}= 
\sigma_{(0)}\frac{\Gamma_{0} B_{W \to \nu e}}{\Gamma}  
+ \as \sigma_{(1)}\frac{\Gamma_{0}B_{W \to \nu e}}{\Gamma}
+\sigma_{(0)}\frac{\as \Gamma_{1} B_{W \to \nu e}}{\Gamma}
\end{equation}
The difference between these two expressions is of $O(\as^2)$ and is given by
\beqn \label{eq:diff}
\sigma B_{t \to b \nu e+X}-\sigma B_{t \to b \nu e} &=&
\as \sigma_{(1)}\frac{\Gamma_{0}B_{W \to \nu e}}{\Gamma}
- \as \sigma_{(1)} B_{W \to \nu e} \nn \\
&=& -\as^2 \sigma_{(1)} \frac{\Gamma_{1}B_{W \to \nu e}}{\Gamma}
\eeqn
\begin{table*}[tb]
\caption{\small Comparison of LO and NLO cross sections for single $t$-quark
production at the Tevatron and LHC. The NLO calculation is performed both
without including QCD effects in the decay ($\sigma B_{t \to b \nu e}$) and
also when it is included ($\sigma B_{t \to b \nu e+X}$).
The top quark mass is $m_t=178$ GeV and cross sections are 
evaluated using MRST2002 NLO PDFs with all scales set to
$m_t$. The errors represent Monte Carlo statistics only.
Note that the values of $\Gamma_t$ at LO and NLO are $1.6511$~GeV and
$1.5077$~GeV respectively and the branching ratio of the $W$ into
leptons is ${\rm Br}(W \to e \nu) = 0.1104$.
The differences in the rates are in good agreement with the formulae in
Eq.~(\ref{eq:diff}). 
\label{tab:ratecomp}}
\begin{center}
\begin{ruledtabular}
\begin{tabular}{lllll}
Process  & $\sqrt{s}$~[TeV] &
$\sigma_0 B_{t \to b \nu e}$ (fb) &
$\sigma B_{t \to b \nu e}$ (fb) &
$\sigma B_{t \to b \nu e+X}$ (fb) \\
\hline
$s$-channel, $p\bar p$ ($t$) &1.96 & 31.54 $\pm$0.01  & 44.64 $\pm$0.03
& 45.88  $\pm$0.03 \\
$s$-channel, $pp$ ($t$)      &14   & 503.9 $\pm$0.2   & 681.0 $\pm$0.4
& 698.7  $\pm$0.4 \\
\hline
$t$-channel, $p\bar p$ ($t$) &1.96  & 111.34 $\pm$0.06 & 113.95  $\pm$0.12
& 113.96  $\pm$0.12 \\
$t$-channel, $pp$ ($t$)     &14    & 17690.  $\pm$8    &    17048.
$\pm$16  & 16975.  $\pm$16. \\
\end{tabular}
\end{ruledtabular}
\end{center}
\end{table*}
As shown in Table~\ref{tab:ratecomp}
the numerical differences are less than $3\%$ for the $s$-channel process
and less than $0.5\%$ for the $t$-channel process.

\subsection{Signals and Backgrounds at NLO}
We shall consider the signal for single top production to be 
the presence of a lepton, missing energy and two jets,
one of which is tagged as a $b$-jet. In the case where we have two tagged 
jets, we choose the jet to be assigned to the top quark at random.
For clarity, we shall describe the processes at the parton level
choosing specific partons in the initial state. In our program 
we sum over all the species of partons present in the initial proton
and antiproton. For reactions considered at NLO, there
can also be additional partons in the final state. 
We use the Run II $k_T$-clustering
algorithm to find jets, with a pseudo-cone of size $R=1.0$~\cite{Blazey:2000qt}.
The first reactions to consider are the two signal processes
\beq
\dk{u+\bar{d}} {t +\bar{b}}{\nu +e^+ + b}
\label{eq:schannelwdk}
\eeq
and 
\beq
\dk{b+u} {t +d}{\nu +e^+ + b} \; .
\label{eq:tchannelwdk}
\eeq
Note that we present numerical results for the processes shown in 
Eqs.~(\ref{eq:schannelwdk},\ref{eq:tchannelwdk}), summed over species of 
initial partons, i.e. the production of a $t$-quark (rather than a $\bar{t}$)
with the decay $t \to \nu+e^+ + b+X$.  Thus in a hypothetical experiment
with equal perfect acceptances for electrons of both charges and for muons  
of both charges the signal (and background) will be four times bigger,
(if we ignore possible signatures coming from $W\to \tau \nu_\tau$).  
Unless explicitly stated otherwise, our NLO results for the signal processes are
calculated including QCD corrections in both the production and decay of
the top quark.

The background processes which we consider are of several types.
The irreducible backgrounds are,
\beq
\label{eq:wbbchannelwdk}
\dk{u+\bar{d}}{W^+ + b + \bar{b}}{\nu +e^+}
\eeq
\beq
\label{eq:wbchannelwdk}
\dk{u+b}{W^+ +d + b}{\nu +e^+} \; .
\eeq
\beq
\label{eq:wzchannelwdk}
\bothdk{u+\bar{d}}{W^++}{Z}{b + \bar{b}}{\nu +e^+}
\eeq
There are also backgrounds related to $t \bar{t}$ 
production which contribute to the $W+$~2 jets process.
The case where both top quarks decay leptonically,
\beq
\label{eq:ttbarleptchannelwdk}
\bothdk{u+\bar{u}}{t+}{\bar{t}}{\bar{b} + e^- + \bar{\nu}}{\nu +e^+ + b}
\eeq
contributes if the electron, $e^-$ (or muon, $\mu^-$) fails the cuts.
If one of the top quarks instead decays hadronically then there can also
be a contribution when only two jets are observed, either 
because of merging or because the extra jets lie outside the 
acceptance (or both), 
\beq
\label{eq:ttbarhadchannelwdk}
\bothdk{u+\bar{u}}{t+}{\bar{t}}{\bar{b} + q + \bar{q}}{\nu +e^+ + b}
\eeq

A significant background process 
involves $W$ + two light jet production
where one of the light quark jets fakes a $b$-quark,
\beq
\label{eq:w2jetchannelwdk}
\dk{u+\bar{d}}{W^+ + {\rm 2~jets}}{\nu +e^+}
\eeq
Further backgrounds involve the mistagging of a $c$-quark as a $b$-quark,
\beq
\label{eq:wsgoestocchannelwdk}
\dk{u+\bar{s}}{W^+ +u + \bar{c}}{\nu +e^+}
\eeq
\beq
\label{eq:wccchannelwdk}
\dk{u+\bar{d}}{W^+ + c + \bar{c}}{\nu +e^+}
\eeq
\beq
\label{eq:wcchannelwdk}
\dk{u+c}{W^+ +d + c}{\nu +e^+}
\eeq

 Our estimation of the rates for these processes depends on  
the cuts shown in Table~\ref{tab:cuts}, which have been chosen to mimic those
used in an actual Run II analysis~\cite{Juste:2004an}. 
\begin{table}
\caption{Cuts for the single top analysis presented in this section.\label{tab:cuts}}
\begin{center}
\begin{ruledtabular}
\begin{tabular}{ll} 
Lepton~$p_T$     &$p_T^{e} > 20$~GeV \\
\hline
Lepton pseudorapidity  &     $|\eta^{e}| < 1.1$ \\
\hline
Missing $E_T$          &     $\slsh{E_T} > 20$~GeV \\
\hline
Jet $p_T$              &     $p_T^{\mbox{jet}} > 15$~GeV \\
\hline
Jet pseudorapidity     &     $|\eta^{\mbox{jet}}|  < 2.8$  \\
\hline
Mass of $b+l+\nu$      &     $ 140 < m_{bl\nu} < 210$~GeV \\
\end{tabular}
\end{ruledtabular}
\end{center}
\end{table}
As well as the normal jet
and lepton cuts, we perform a cut on the missing transverse energy and
the mass of the `$b+l+\nu$'-system, $m_{bl\nu}$. 
The missing transverse energy vector is the negative of 
the vector sum of the transverse
energy of the observed jets and leptons. The mass of the 
putative top system is determined by reconstructing the $W$ and combining
it with the tagged $b$-jet. In reconstructing the $W$
the longitudinal momentum of the neutrino
is fixed  by constraining the mass of the $e \nu$ system to 
be equal to $\MW$. The two solutions for the longitudinal momentum
of the neutrino are,
\beq
p_L^{\nu} = \frac{1}{2 |p_T^{e}|^2}
\Big[p_L^{e} \; \big(\MW^2-M_T^2+2 |p_T^{e}| |p_T^{\nu}| \big) 
\pm 
E^e \; \sqrt{(\MW^2-M_T^2) (\MW^2-m_T^2+4 |p_T^{e}| |p_T^{\nu}|)}
\Big]
\eeq
In this equation $p_T^{\nu}$ is the measured missing transverse
energy and 
\beq
M_T^2=(|p_T^{e}|+|p_T^{\nu}|)^2- (\vec{p}_T^{\;e}+\vec{p}_T^{\;\nu})^2
\eeq
is the transverse mass of the $W$. 
We resolve the twofold ambiguity in $p_L^{\nu}$ by choosing 
the solution which gives the 
largest (smallest) neutrino rapidity for the $W^+ (W^-)$.

Our results are shown in 
Table~\ref{tab:xsect}.
\begin{table}
\caption{Cross sections for the Tevatron Run II ($p{\bar p}$, $\sqrt s =1.96$~TeV)
in femtobarns\label{tab:xsect}, calculated
with MRST2002 and using the renormalization and factorization scale $\mu$.
The cross sections contain two jets, subject to the cuts
of Table~\ref{tab:cuts}.
Next-to-leading order cross sections are also shown, where 
known. The last column estimates the importance of signal and background processes
by including nominal tagging and mistagging efficiencies, using the 
NLO cross section where possible.}
\begin{center}
\begin{ruledtabular}
\begin{tabular}{llllll} 
Process  & Scale $\mu$ & $\sigma_{LO}$~[fb] & $\sigma_{NLO}$~[fb] & Efficiency
& $\sigma$~[fb] \\
\hline
$s$-channel single top, Eq.~(\ref{eq:schannelwdk}) &   $m_t$ & 10.3 & 11.7 & $1-(1-\varepsilon_b)^2$ & 7.4 \\
$s$-channel (with decay radiation) &   $m_t$ & 10.3 & 11.3 & $1-(1-\varepsilon_b)^2$ & 7.2 \\
\hline
$t$-channel single top, Eq.~(\ref{eq:tchannelwdk}) &   $m_t$ & 38.8 & 29.4 & $\varepsilon_b$ &  11.8  \\
$t$-channel (with decay radiation) &   $m_t$ & 38.8 & 26.6 & $\varepsilon_b$ &  10.6  \\
\hline
$Wb \bar{b}$, Eq.~(\ref{eq:wbbchannelwdk}) &  $\MW$ &  36.0 & 47.5 & $1-(1-\varepsilon_b)^2$  & 30.4 \\
\hline
$W+bj$, Eq.~(\ref{eq:wbchannelwdk}) & $\MW/4$ & 26.5 & - & $\varepsilon_b$ &  10.6  \\
\hline
$WZ$, Eq.~(\ref{eq:wzchannelwdk}) & $\MW$ & 3.64 & 3.91 &   $1-(1-\varepsilon_b)^2$ & 2.5  \\
\hline
$t \bar{t}$, Eq.~(\ref{eq:ttbarleptchannelwdk}) &$m_t$ &  4.34 & - & $ 2 \times(1-(1-\varepsilon_b)^2)$ & 5.6  \\
\hline 
$t \bar{t}$, Eq.~(\ref{eq:ttbarhadchannelwdk}) & $m_t$ & 4.94 & -  &$1-(1-\varepsilon_b)^2$ & 3.2  \\
\hline
$W$+2 jet, Eq.~(\ref{eq:w2jetchannelwdk}) &  $\MW$ & 5530. & 7030. &  $1-(1-f_J)^2$  &  35.1  \\
\hline
$u\bar{s} \to Wu \bar{c}$, Eq.~(\ref{eq:wsgoestocchannelwdk})  &  $\MW$ & 324.&-  & $f_c $ & 19.4 \\
\hline
$Wc \bar{c}$, Eq.~(\ref{eq:wccchannelwdk}) &  $\MW$ & 36.0 & 47.5 &      $1-(1-f_c )^2$  & 5.5 \\
\hline
$W+cj$, Eq.~(\ref{eq:wcchannelwdk})   & $\MW/4$ & 54.7 & - &$f_c $  &  3.3   \\
\end{tabular}
\end{ruledtabular}
\end{center}
\end{table}
In this table we first show the cross sections calculated assuming a
perfectly efficient detector, both at LO and NLO where possible. The
column labeled `Efficiency' gives the rescaling factor that should be
applied in order to take account of experimental efficiencies and fake
rates. To give some idea of the importance of the background
processes, we have used the nominal (and perhaps optimistic) values of
these quantities shown in Table~\ref{tab:eff} to obtain the final
column of Table~\ref{tab:xsect}.
\begin{table}
\caption{Nominal efficiencies used throughout this section.\label{tab:eff}}
\begin{center}
\begin{ruledtabular}
\begin{tabular}{ll} 
$b$-tagging efficiency &     $\varepsilon_b =40\%$ \\
\hline 
$c$-mistagged as $b$   &     $f_c= 6\%$ \\
\hline
Jet fake rate          &     $ f_J=0.25\%$ \\
\end{tabular}
\end{ruledtabular}
\end{center}
\end{table}

\begin{figure}
\begin{center}
\includegraphics[,angle=270,width=\columnwidth]{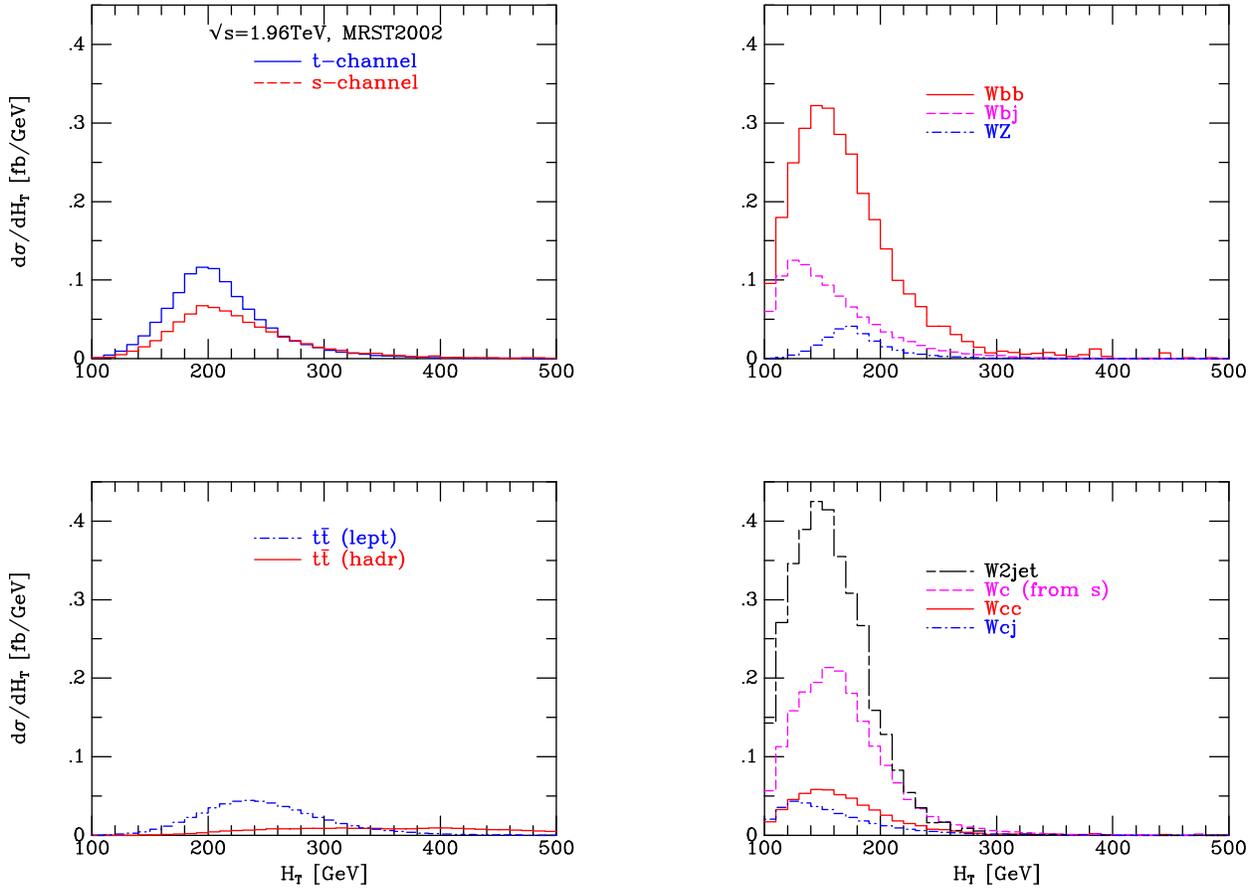}
\caption{
Predictions for $H_T$ showing single top signal, (top left), 
irreducible backgrounds (top right), 
$t \bar{t}$~backgrounds, (bottom left) and jet
and charm fake rate (bottom right)\label{fig:ht}.
Next-to-leading order estimates are used for the cross sections 
where available.}
\end{center}
\end{figure}

\begin{figure}
\begin{center}
\includegraphics[,angle=270,width=\columnwidth]{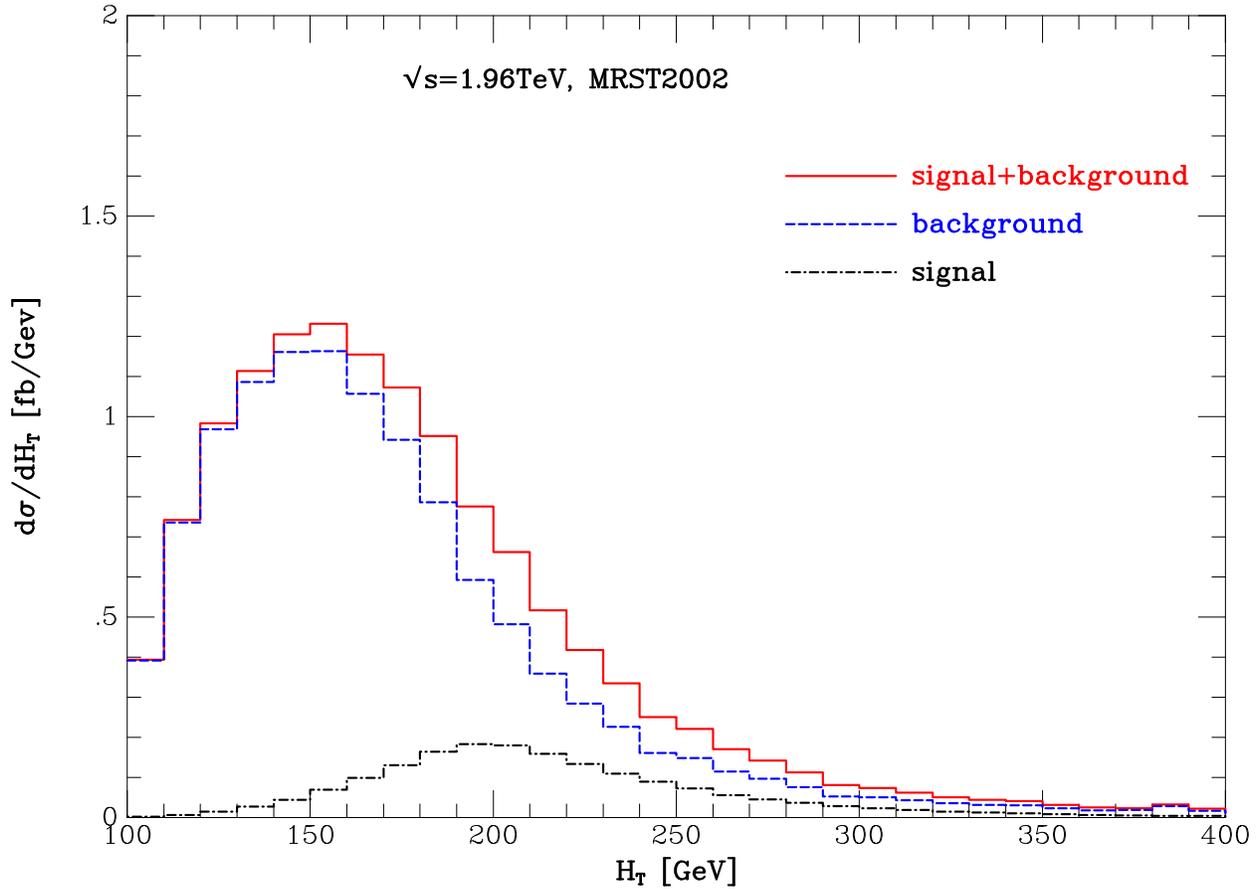}
\caption{The $H_T$ distributions of signal, background and signal plus background. The curves
correspond to the sum of the appropriate distributions shown in Figure~\ref{fig:ht}.
\label{fig:htoverall}}
\end{center}
\end{figure}

A strategy for searching for the single top processes at the Tevatron involves examining
the $H_T$ distribution~\cite{Juste:2004an}, where $H_T$ is defined by,
\beq
H_T=|p_T(\mbox{lepton})|+|\slsh{E}_T|+\sum |p_T(\mbox{jet})| \; .
\eeq
In Fig.~\ref{fig:ht} we show the $H_T$ distributions of the signal and background
processes described above, rescaled by the nominal efficiencies given 
in Table~\ref{tab:eff}. As in the table, the NLO calculation is used wherever it is known
and the signal processes include gluon radiation in the decay of the top quark.
We see that the $H_T$ distribution is harder
for the single top distributions, than all backgrounds, except those 
which involve top production (which are however, small).
Despite this fact, one sees that the background rates are still large
in the region of $H_T$ where the signal processes peak. This is
demonstrated more clearly in Figure~\ref{fig:htoverall}, where we show
the $H_T$ distribution for the sum of all the backgrounds, the sum of
the two single top processes and the total when combining signal and
background. Although the single top signal represents about $50\%$ of
the events in the bins of interest, an observation of this top
production mechanism using the $H_T$ distribution is heavily reliant
on accurate predictions of both the rates and shapes of the background
processes. We stress that Figs.~\ref{fig:ht} and~\ref{fig:htoverall}
depend on the particular values chosen for the efficiencies and will
improve if better values are achieved.

\begin{figure}
\begin{center}
\includegraphics[,angle=270,width=\columnwidth]{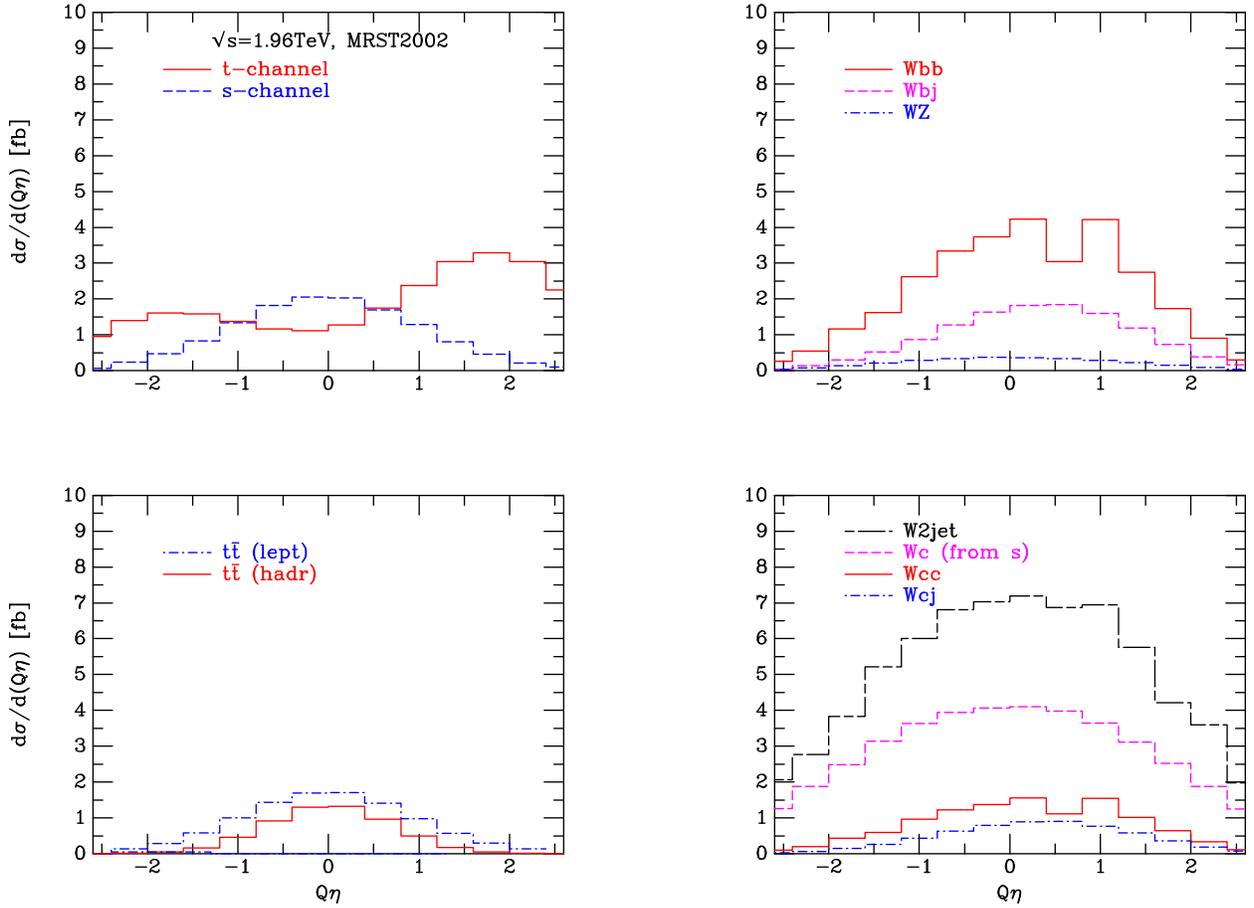}
\caption{Predictions for $Q \eta$ showing single top signal, (top left), 
irreducible backgrounds (top right), $t \bar{t}$~backgrounds, (bottom
left) and jet and charm fake rate (bottom right)\label{fig:eta}.
Next-to-leading order estimates are used for the cross sections where
available.}
\end{center}
\end{figure}

\begin{figure}
\begin{center}
\includegraphics[,angle=270,width=\columnwidth]{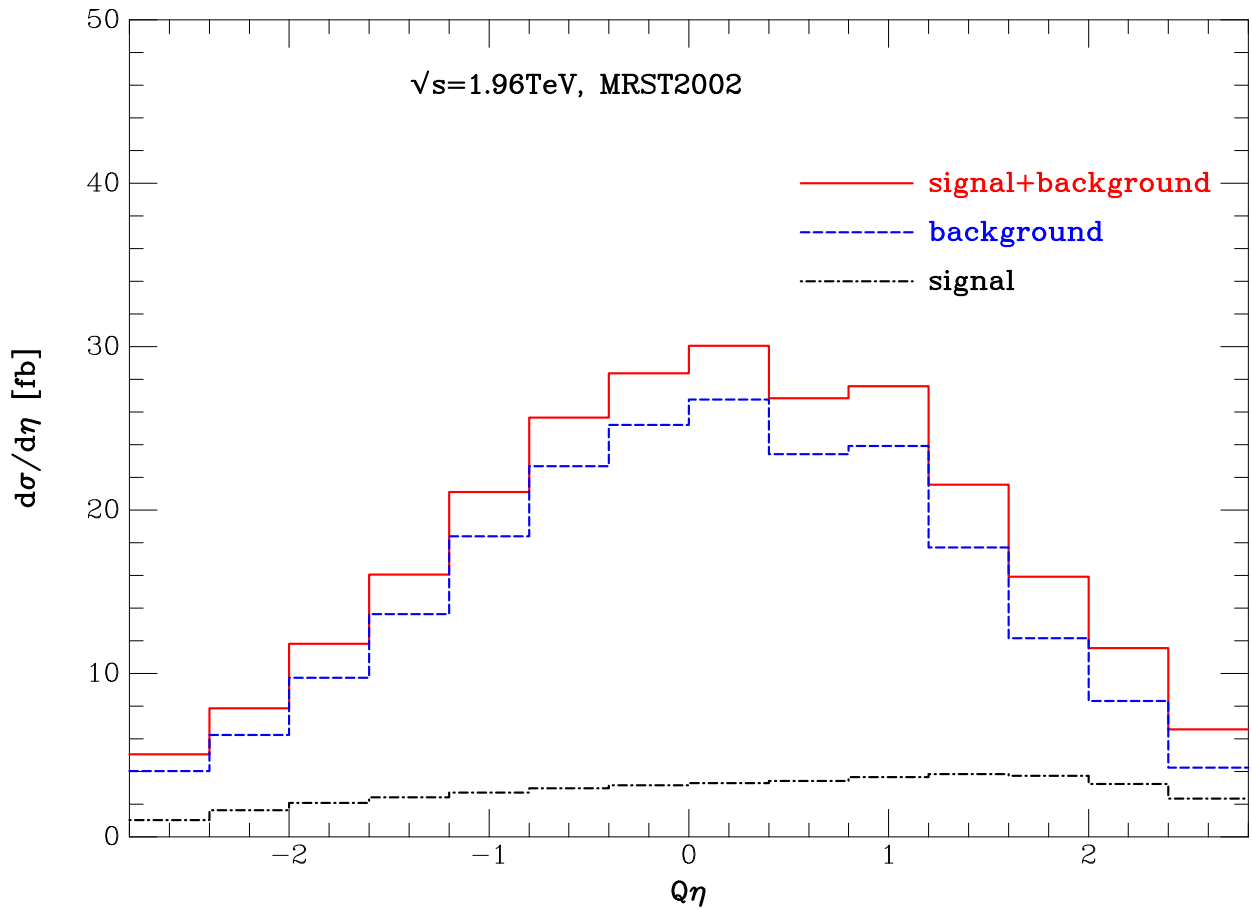}
\caption{The $Q\eta$ distributions of signal, background and signal plus background. The curves
correspond to the sum of the appropriate distributions shown in Figure~\ref{fig:eta}.
\label{fig:etaoverall}}
\end{center}
\end{figure}

A further quantity that may be used to discriminate between the $t$-channel signal process and
backgrounds is $Q\eta$, where $Q$ is the charge of the lepton (in units of the positron charge) and
$\eta$ is the pseudorapidity of the untagged jet. In this analysis~\cite{Juste:2004an} an
additional cut is applied, requiring that one of the two jets must have a $p_T> 30$~GeV. This
extra  cut serves to reduce the dominant backgrounds in
Table~\ref{tab:xsect} by $15$--$30\%$ whilst leaving the signal virtually
unchanged. Since this is a targeted search for the $t$-channel process
we have also rejected events with two tagged jets, since at leading order,
Eq.(\ref{eq:tchannelwdk}) contains only one $b$-jet. 
Results for the $Q\eta$ distribution are shown in Fig.~\ref{fig:eta}, demonstrating
that the $t$-channel process is enhanced at large pseudorapidity compared to the backgrounds.
As before, even the pronounced
effect observed in this distribution can be hidden when comparing the signal to the
sum of all backgrounds, as shown in Fig.~\ref{fig:etaoverall}.

\begin{figure}
\begin{center}
\includegraphics[angle=270,width=\columnwidth]{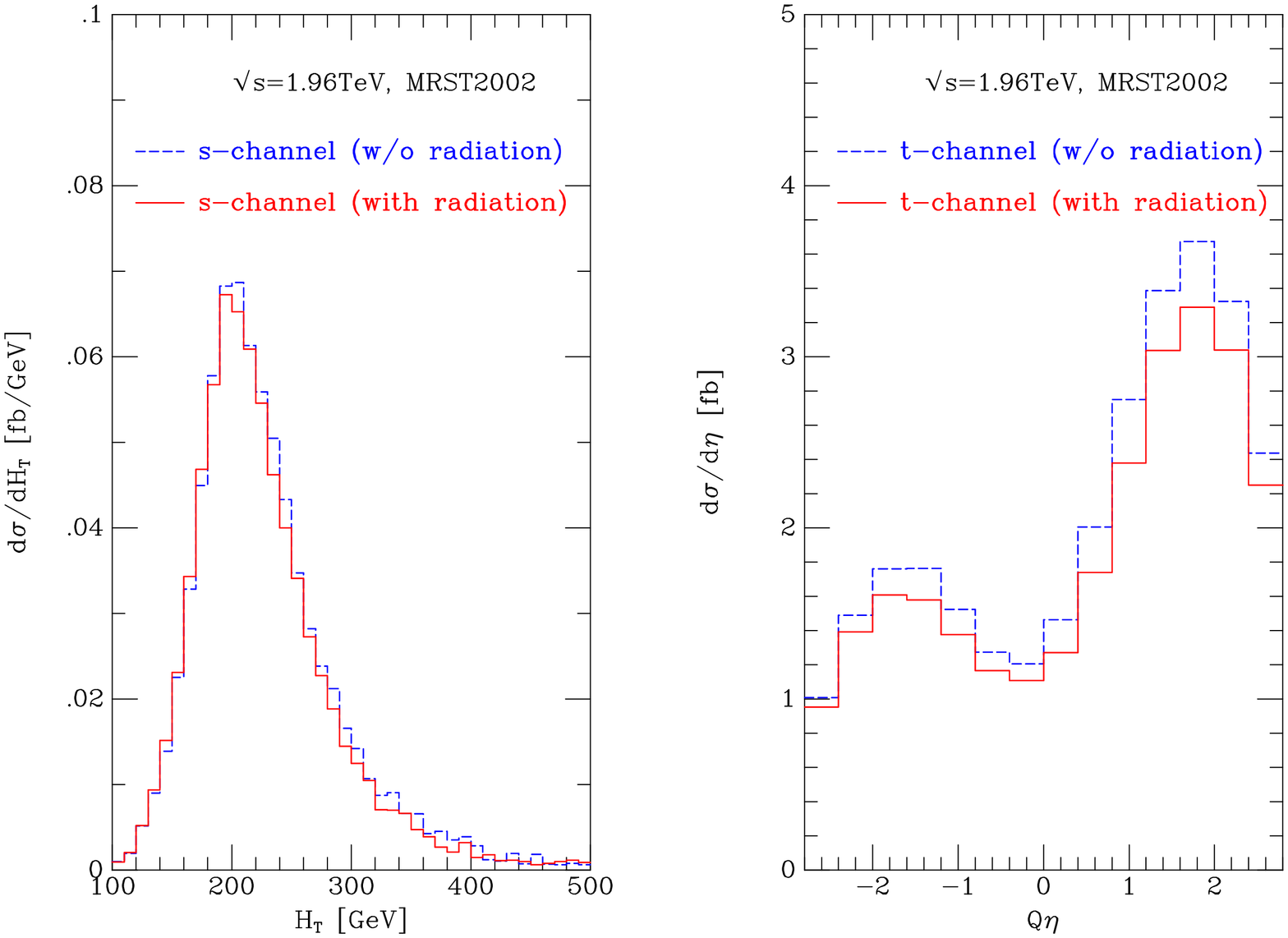}
\caption{Comparison of the NLO distributions for $H_T$ in the $s$-channel process
(left) and $Q\eta$ in the $t$-channel process (right), 
with and without radiation in decay. \label{fig:comparedk}}
\end{center}
\end{figure}

Finally, we end this section with a comment on the effect of including the radiation
in the top quark decay at NLO. We find that although the overall rate is lowered
(cf. Table~\ref{tab:xsect}), the shapes of the $H_T$ and $Q\eta$ distributions are
not altered significantly. This is demonstrated in Fig.~\ref{fig:comparedk}, where we
compare the $H_T$ ($s$-channel) and $Q\eta$ ($t$-channel) distributions  with and
without radiation in the decay. However this is a feature of our specific
choice of cuts and we do not know this to be true in general.


\section{Conclusions}
We have presented the first results of our NLO Monte Carlo program
which describes the signal and background for single top production.
The new feature of our analysis is the inclusion of the decay of the
top quark. Since our analysis is performed at next-to-leading order,
we have included radiative effects both in the production and 
in the decay. Radiation in the decay is performed using a new subtraction
method, which may be useful in other contexts.

We have studied the effects of the NLO corrections on two
distributions that are relevant for single top searches at the
Tevatron. The inclusion of the top quark decay is imperative for such
an analysis for two reasons. Firstly, the cuts that are applied (in
order to match the experimental analysis) are on the decay products
and not on the top quark itself. Second, one of the distributions of
interest, $H_T$, can only be calculated when the momenta of the decay
products are known. We have also performed the calculation of a number
of important backgrounds at NLO and examined the feasibility of using
variables such as $H_T$ and $Q\eta$ to discriminate between the single
top signal and the main backgrounds. We find that the inclusion of
radiation in the decay makes very little difference in the shapes of
these distributions, but further decreases the exclusive two-jet
signal cross section.  Our treatment is obviously approximate, because
we assume $p_T$-independent efficiencies, stable $b$ and $c$ quarks
and include no showering or hadronization. Nevertheless it confirms
that the search for single top is extremely challenging and suggests
that significantly larger tagging efficiencies and/or methods with
greater discriminatory power will be needed to observe single top
production. We hope to have contributed to this search by providing
more reliable information on the kinematic structure of both the
signal and the background.

\begin{acknowledgments}
R.K.E would like to acknowledge helpful discussions with W.A. Bardeen.
This work has been supported by the Universities Research Association, Inc. 
under contract No. DE-AC02-76CH03000 with the U.S. Department of Energy
and also by the U.S. Department of Energy under Contract No. W-31-109-ENG-38.
\end{acknowledgments}

\appendix
\section{Integration of dipoles}
\label{app:alphadep}
As described in the text of the paper the production cross section 
was calculated using a subtraction method~\cite{Ellis:1980wv}. The particular 
subtraction terms were taken from ref.~\cite{Catani:1997vz} (CS)
for the subtraction terms with massless particles 
and from ref.~\cite{Catani:2002hc} (CDST) for the 
dipoles involving massive particles in the final state.

In this appendix we describe two minor modifications of
the methods in those references. The first is that we allow ourselves
the freedom to work in the four-dimensional helicity scheme. This scheme
has been used for the calculation of many virtual matrix elements, so we 
believe that our results may be useful in other contexts.

The second modification is that, following Nagy and
Tr\'ocs\'anyi~\cite{Nagy:1998bb,Nagy:2003tz} we introduce a tuneable
parameter $\alpha$ which can be used to reduce the range of
integration over the singular variable. $\alpha=1$ corresponds to
integration over the whole dipole phase space and for $\alpha<1$ the
volume of phase space for the dipole subtraction is reduced.  As
detailed in~\cite{Nagy:2003tz} this serves a number of purposes.
Since the subtraction is smaller, the mis-match between real matrix
elements and subtractions is also reduced, leading to fewer events
where the event and counter-event fall into different bins. In
addition, for small $\alpha$ the counter-terms are only calculated in
events where they actually play a role in canceling a
singularity. This can lead to a saving in computer time. Lastly, the
lack of $\alpha$-dependence in the results is a valuable check of the
numerical implementation.  For the massless case the integrations of
the dipoles in reduced phase space volumes have been provided
in~\cite{Nagy:2003tz}.  The results for the $\alpha$ dependence of the
massive dipoles are new.

To keep the appendix short we try and use a similar notation to 
refs.~\cite{Catani:1997vz,Catani:2002hc} and refer back to specific 
equations in those references 
(CS~\cite{Catani:1997vz} and CDST~\cite{Catani:2002hc})
when appropriate.

\subsection{Initial-state emitter with initial state spectator}
As explained above we have used a slight generalization of the dipole
phase space (CS, Eq.~(5.151)) where the variable ${\tilde v_i}$ 
is modified by the factor $\Theta(\alpha-{\tilde v}_i)$.
The variable ${\tilde v_i}$ is the rescaled value of the propagator, defined as
\beq   
{\tilde v_i}= \frac{p_a p_i}{p_a p_b}
\eeq   
where $p_a$ is the initial state emitter, $p_i$ is the emitted parton and 
$p_b$ is the other initial state parton which is the spectator. 
For a full description see section 5.5 of CS.
The dipole integrand which we subtract is obtained by introducing $\eta$
into CS, Eq.~(5.145) for the $q,q$ case 
and CS, Eq.~(5.146) for the $g,q$ case:
\beq
\bra{s}
{\bom V}^{q_ag_i,b}(x_{i,ab}) \ket{s'}
= 8\pi \mu^{2\ep} \as\; C_F\;\delta_{ss'} \left[ \frac{2}{1-x_{i,ab}}
- \frac{}{} (1+ x_{i,ab}) - \eta \ep (1-x_{i,ab})) \right]
\;,
\eeq
\beqn
\bra{s}
{\bom V}^{g_a{\bar q}_i,b}(x_{i,ab}) \ket{s'}
= 8\pi \mu^{2\ep} \as\; T_R\;
\left[ 1 - \eta \ep -2 x_{i,ab}(1-x_{i,ab}) \right]
\;\delta_{ss'} \;,
\eeqn
As in Eq.~(\ref{eq:etadef})
for $\eta=1$ we are in the 't Hooft-Veltman scheme. 
Setting $\eta=0$ we are in the 4-dimensional helicity scheme.

The result for the $q,q$ case 
is a generalization of CS, Eq.~(5.155) and is given by,
\beqn
&&{\tilde \cV}^{q,q}(x;\ep,\alpha)
=C_F \Bigg\{\Bigg(\frac{1}{\ep^2}-\frac{\pi^2}{6}\Bigg) 
  \delta(1-x) \nn \\
&+& \eta(1-x)-(1+x) \Big(2 \ln(1-x)-\frac{1}{\ep}\Big)
\nn \\
&+&\Theta(1-x-\alpha) \frac{(1+x^2)}{(1-x)} 
 \ln\Big(\frac{\alpha}{1-x}\Big) 
\nn \\
      &-& \frac{2}{\ep} \frac{1}{\big[1-x\big]_{+}}
  +4 \Bigg[\frac{\ln(1-x)}{1-x}\Bigg]_{+} \Bigg\} +\Oe{}\:
\eeqn
The result for the $g,q$ case 
(after averaging in $d$ dimensions in the 't Hooft Veltman scheme)
is given by,
\beqn
&&{\tilde \cV}^{g,q}(x;\ep,\alpha)
=T_R \Bigg\{\Big((1-x)^2+x^2\Big) \nn \\
&\times & \Bigg[2 \ln(1-x)-\frac{1}{\ep}
    +\Theta(1-x-\alpha) \ln\bigg(\frac{\alpha}{1-x}\bigg)\Bigg] \nn \\
  &+&2 \eta x (1-x) \Bigg\} +\Oe{}\:
\eeqn
In the limit $\alpha=1$ (and $\eta=1$)
these functions reduce to those derived 
by the explicit expansion of CS, Eq.~(5.155).
They are also in agreement with the results of Nagy~\cite{Nagy:privcomm}.

In the above equations we have followed the conventions of CS. However in CS the overall
factor multiplying the functions ${\tilde \cV}$ contains a term of the form,
(cf. CS Eq.~(5.152))
\beq
\Bigg( \frac{4 \pi \mu^2}{2 p_a  p_b} \Bigg)^\epsilon \; .
\eeq
In our program we choose to write this factor as         
\beq
\Bigg( \frac{4 \pi \mu^2}{2 p_a  p_b} \Bigg)^\epsilon =
\Bigg( \frac{4 \pi \mu^2 x }{2 {\tilde p_{ai}}  p_b} \Bigg)^\epsilon 
\approx \Bigg( \frac{4 \pi \mu^2}{2 {\tilde p_{ai}}  p_b} \Bigg)^\epsilon 
\Big(1+\ep \ln x  + \ldots \Big) \; ,
\eeq
since it is the transformed momenta which are held fixed when we perform the $x$ integration. In
our program the additional $\ln x$ terms are  included in our results for the integrated dipoles.
These additional terms containing $\ln x$ are accounted for in the paper of CS in a different
fashion. For a similar reason, our program contains additional $\ln x$ terms in the numerical
implementation of Eqs.~(\ref{ifinteg},\ref{ifintegb}).

\subsection{Initial-state emitter with final-state spectator}

We have two cases which we have to consider
\beqn
&& \mbox{a) Initial } q  \to q+g, \mbox{with a massive final state spectator}; \nn \\
&& \mbox{b) Initial } q  \to q+g, \mbox{with a massless final state spectator} \nn
\eeqn

For case (a), the phase space is the generalization of CDST, Eq.~(5.79) with an
extra factor of $\Theta(\alpha-z_i)$ which implements the reduction in the phase
space volume. The variable $z_+$ is defined by,
\beq
z_+ = \frac{1-x}{1-x+\mu_Q^2}.
\eeq
The dipole integrand is given by a generalization of 
CDST, Eq.~(5.81). The result is
\def\mut{\tilde \mu}
\beqn
\label{ifinteg}
&&I^{qq}_Q(x;\ep,\alpha) = C_F\Bigg\{ -\frac{1}{\ep} 
 \Big[ \frac{2}{[1-x]_{+}}-1-x\Big]
 + \delta(1-x) \Bigg[ \frac{1}{\ep^2}+\frac{\pi^2}{6}
   + \frac{1}{\ep}\ln(1+\mut^2) \nn \\
 &+& 2 \Li_{2}(-\mut^2)
 +2 \ln(\mut^2) \ln(1+\mut^2)- \frac{1}{2} \ln^2(1+\mut^2) \Bigg] \nn \\
 &+&4\;  \Big[\frac{\ln(1-x)}{(1-x)}\Big]_{+}
 -2\;  \frac{\ln(1+{\tilde \mu}^2)}{[1-x]_{+}} 
 -\frac{2 }{(1-x)} \ln\Big(\frac{2-x+x\mut^2}{1+\mut^2}\Big)
\nn \\
 &-& \Theta(z_+-\alpha) \Bigg[ \left( \frac{2}{1-x} \right)
  \log \left( \frac{z_+ (1-x+\alpha )}{\alpha(1-x+z_+)} \right)
  -(1+x) \ln \left( \frac{z_+}{\alpha} \right) \Bigg]
\nn \\
 &-&  (1+x)\ln\Big(\frac{(1-x)^2}{(1-x+x\mut^2)}\Big) + \eta (1-x)
  \Bigg\} +\Oe{}\:
\eeqn
For $\alpha=1$ and $\eta=1$ this agrees with CDST, Eqs.~(5.90) and (5.88).
Here we have introduced the variable 
\beq
{\tilde \mu^2} = \frac{\mu^2}{x} = \frac{m^2}{2 {\tilde p_{ai}} {\tilde p_j}}
\eeq
which only depends on ${\tilde p_{ai}}$ and  ${\tilde p_j}$ which are defined in
CDST Eq.~(5.73).
This is helpful since it is these transformed momenta which are held fixed
when we do the $x$ integration.

For case (b), the dipole phase space is given by CS, Eq.~(5.72) with an additional
factor of $\Theta(\alpha-u_i)$. The integrand is given by the
usual $\eta$-dependent generalization of CS, Eq.~(5.77).
Performing the integration yields,
\beqn \label{ifintegb}
\cV^{q,q}(x;\ep,\alpha) &=& C_F \Bigg\{
   \eta(1 - x) - \frac{2}{1-x}\ln\Big(\frac{1+\alpha-x}{\alpha}\Big) 
  + \Big(\frac{1}{\ep}-\ln(\alpha)-\ln(1-x)\Big) (1+x) \nn \\
  &+& \delta(1-x)\Big(\frac{1}{\ep^2} +  \frac{\pi^2}{6} \Big) 
  + 4 \Big[\frac{\ln(1-x)}{(1-x)}\Big]_{+}
  -\frac{2}{\ep} \Big[\frac{1}{1-x}\Big]_{+} \Bigg\}+\Oe{}\:
\eeqn
This result for $\eta=1$ is in agreement with CS, Eq.~(5.83). 
It is also in agreement 
with the $\alpha \neq 1$ results of Nagy~\cite{Nagy:privcomm}.
Since the $\mu \to 0$ limit is smooth it can also be obtained from 
the limit of Eq.~(\ref{ifinteg}).

\subsection{Final-state emitter with initial-state spectator}

Once again there are two cases which we have to consider,
\beqn
&& \mbox{a) Final } Q  \to Q+g, \mbox{with a massless initial-state spectator;} \nn \\
&& \mbox{b) Final } q  \to q+g, \mbox{with a massless initial-state spectator.} \nn
\eeqn

We shall deal with the two cases in turn.
The phase space for the first case is given by
CDST, Eq.~(5.48) with the addition of the factor $\Theta(x-1+\alpha)$ to
reduce the phase-space volume. The integrand is given by CDST, Eq.~(5.50) and yields a
result which we choose to decompose into three separate contributions.
Our decomposition of $I_{ij}^a$ is into a delta-function, plus-distribution and
regular parts.
\beq
\label{eq:I_gQa}
I_{gQ}^a(x;\ep,\alpha)  = C_F \left\{
\delta(1-x) \, J_{gQ}^{a\; \rm\delta}(\mu_Q,\ep,\alpha)
+J_{gQ}^{a \; +}(x, \mu_Q,\alpha)
+ J_{gQ}^{a\;\rm R}(x,\mu_Q,\alpha) \right\}
+\Oe{}, \nn \\
\eeq
where we have introduced the variable (c.f. Eq.~(5.45) of CDST)
\beq
\mu_Q^2 = \frac{m^2}{2 {\tilde p_{ij}} {p_a}}.
\eeq
This is a different choice than the one given in 
Eq.~(5.56) of ref.~\cite{Catani:2002hc} and more suitable for implementation
into our program. The three contributions to Eq.~(\ref{eq:I_gQa}) are 
\beqn
J_{gQ}^{a \; \delta}(\mu, \ep,\alpha)& = &\frac{1}{\ep}
-\frac{1}{\ep} \ln\Big(\frac{1+\mut^2}{\mut^2}\Big)
-2 \Li_2(-\mut^2) -\frac{\pi^2}{3}+2 
+2 \ln (\alpha) \Big[\ln\Big(\frac{1+\mut^2}{\mut^2}\Big) - 1 \Big] \nn \\
&+&\frac{1}{2}\ln^2\mut^2
+\frac{1}{2}\ln^2(1+\mut^2)-2 \ln \mut^2 \ln (1+\mut^2)
+\ln (\mut^2)
\eeqn
\beq
J_{gQ}^{a \;\rm +}(x,\mu,\alpha) =
\Big[\ln\Big(\frac{1+{\tilde \mu}^2}{{\tilde \mu}^2}\Big) - 1 \Big]
 \left(\frac{2}{1-x}\right)_{1-\alpha} \nn \\
\eeq
\beq
[J_{gQ}^{a\; R}(x, \mu,\alpha)] = \Bigg\{
\frac{1-x}{2(1-x+\mu^2)^2} 
+\frac{2}{(1-x)}
\ln\Bigg(\frac{(2-x+\mu^2) \; {\tilde \mu}^2}
             {(1+{\tilde \mu}^2)(1-x+\mu^2)}\Bigg)
  \Bigg\}\; \Theta(x-1+\alpha)
\label{Jdef}
\eeq
where we have defined $\mut^2 = \mu^2/x$ because 
the $x$ integration is performed at fixed  $\mut^2$.
The distribution is interpreted as follows
\beq
\int_0^1 \; dx \; f(x) \Bigg(\frac{1}{1-x}\Bigg)_{1-\alpha} =
\int^1_{1-\alpha} \; dx \; \frac{(f(x)-f(1))}{1-x}
\eeq
so that as $\alpha \to 1$ we recover the normal plus distribution. 

Eq.~(\ref{Jdef}) with $\alpha=1$ should be compared with 
Eqs.~(5.58, 5.59 and 5.60) of CDST.
The comparison is mostly easily performed by showing that the two forms
are identical for $x \neq 1$ and that the integral
\beq \int_0^1 \; dx \; I_{gQ}^a(x;\ep) 
\eeq
is identical for the two forms. 

Now we consider case b), in which both the emitter and spectator quarks
are massless. The phase space is given by CS, Eq.~(5.48) 
and the subtraction is given by the $\eta$-dependent generalization
of CS, Eq.~(5.49). The result is
\beqn
&&\cV_{qg}(x;\ep,\alpha) = C_F \Bigg\{ 
\delta(1-x) \Bigg( \frac{7}{2} -\frac{1-\eta}{2} - \frac{\pi^2}{2} 
 - \frac{3}{2}\ln(\alpha) 
- \ln^2(\alpha) + \frac{3}{2 \ep} + \frac{1}{\ep^2} \Bigg) 
\nn \\
 &+&\frac{2\; \ln(2-x)}{1-x} \; \Theta(x-1+\alpha)
 - 2\Bigg(\frac{\ln(1-x)}{1-x}\Bigg)_{1-\alpha} 
 - \frac{3}{2} \Bigg(\frac{1}{1-x}\Bigg)_{1-\alpha} \Bigg\}+\Oe{}\:
\label{vqgnomass}
\eeqn
Eq.~(\ref{vqgnomass}) is in agreement with CS, Eq.~(5.57) when $\eta=1$. 
It is also in agreement with the $\alpha$-dependent results of
Nagy~\cite{Nagy:privcomm}.

\subsection{Final-state emitter with final-state spectator}
\def\YP{y_{+}}
\def\xp{x_{+}}
We now consider final state radiation of a gluon from a quark line
with a final state spectator.
This can either be radiation off a massive line,
with a massless spectator, or radiation off a massless line with a 
massive spectator: 
\beqn
&& \mbox{a) Final } Q  \to Q+g, \mbox{with massless final-state spectator;} \nn \\
&& \mbox{b) Final } q  \to q+g, \mbox{with massive final-state spectator.} \nn
\eeqn

For case (a) we have the phase space from CDST, Eq.~(5.11) multiplied by
$\Theta(\alpha-\yijk)$ where the $\yijk$ and $z_i$ integrations range over,
\beqn
0 < & \yijk & < 1 \nn \\ 
0 < & \zi & < \frac{(1-\mu_Q^2)\yijk}{[\mu_Q^2+(1-\mu_Q^2)\yijk]}\;.
\eeqn

The dipole 
subtraction is a generalization of CDST, Eq.~(5.16).
\beqn
\label{eq:V_gQk}
\langle s|\bV_{g Q,q}|s'\rangle && =
8\pi\mu^{2\ep}\alpha_S\CF \left\{
\frac{2}{1-\zj(1-\yijk)}
-\frac{\tvijk}{\vijk}\left[(1+\zj) \phi +\frac{m_Q^2}{p_i p_j}\right]
\right\} \delta_{ss'}
\label{dipffsub}
\eeqn

$\tvijk$, $\vijk$ and $\zj=1-\zi$ are defined in CDST, Eqs.~(5.8), (5.14) and
(5.12).  Following CDST, Eq.~(5.23) we divide the dipole subtraction 
into an eikonal piece and a collinear piece.
\beq \label{eq:eikdiv}
I_{gQ,k}(\mu_j,\mu_k;\ep) = C_F \Big[2 I^{\eik}(\mu_j,\mu_k;\ep) 
+I^{\coll}(\mu_j,\mu_k;\ep) \Big]
\eeq 
For a massive quark there is no collinear divergence and hence only the terms
with soft singularities are necessary.
The terms which subtract a putative mass singularity ensure continuity in the 
small mass limit. For a large enough mass they are irrelevant and
they can be removed by setting $\phi=0$. 
 
The eikonal integral is defined by
\beqn
\label{eq:Ieik}
&&\frac{\alpha_S}{2\pi}\frac{1}{\Gamma(1-\ep)}
\biggl(\frac{4\pi\mu^2}{Q^2}\biggr)^\ep I^{\eik}(\mu_Q,0;\ep) = \nn \\
&&\int [\rd p_i(\tpij,\tpk)] \, \frac{1}{2p_i p_j} \,
\frac{8\pi\mu^{2\ep}\alpha_S}{1-\zj(1-\yijk)} 
\Theta(\alpha-\yijk), \nn \\
\eeqn
and performing the integration we obtain,
\beqn
I^{\eik}(\mu_Q,0;\ep) &=&
\frac{\ln(\mu_Q^2)}{2 \ep}
-2 \Li_2(1-\mu_Q^2)
-\ln(\mu_Q^2) \ln(1-\mu_Q^2)-\frac{1}{4}\ln^2(\mu_Q^2) \nn \\
&-&\ln(\alpha) \ln(\mu_Q^2)
-\Li_2\Big(\frac{\mu_Q^2-1}{\mu_Q^2}\Big)
+\Li_2\Big(\frac{\alpha (\mu_Q^2-1)}{\mu_Q^2}\Big)+\Oe{}\:.
\eeqn
Taking $\alpha=1$ we can compare with the appropriate limit of CDST, Eq.~(A.1).

The remaining terms in Eq.~(\ref{dipffsub}) are referred to as collinear subtraction
terms (even though the term proportional to $m_Q^2$ in Eq.~(\ref{eq:V_gQk}) has only 
a soft singularity). The result after integration is,
\beqn
I^{\coll}(\mu_Q,0;\ep)&=& \frac{1}{\ep}+\phi+2
+\ln(\mu_Q^2) (1+\frac{\phi}{2})
+\ln(\mu_Q^2) \frac{(\phi-2)}{1-\mu_Q^2}-2 \ln(1-\mu_Q^2) \nn \\
&+&\frac{\phi}{2}\Big[3 \alpha-2 -\frac{(3-\mu_Q^2)}{(1-\mu_Q^2)} \ln(\alpha+(1-\alpha) \mu_Q^2)
 -\frac{\alpha}{(\alpha+(1-\alpha) \mu_Q^2)}
\Big] \nn \\
 &-&2 \ln(\alpha)+2 \frac{\ln(\alpha+(1-\alpha) \mu_Q^2)}{(1-\mu_Q^2)}+\Oe{}\:.
\eeqn
Setting $\alpha=1$ and $\phi=1$ we recover
\beq
I^{\coll}(\mu_Q,0;\ep)= \frac{1}{\ep}+3
+\frac{3}{2} \ln(\mu_Q^2) 
-\frac{\ln(\mu_Q^2) }{1-\mu_Q^2}-2 \ln(1-\mu_Q^2) 
\eeq
This last result can be extracted from the appropriate limit of
CDST, Eq.~(5.35).

For case (b) the phase space is again taken from CDST, Eq.~(5.11) but
with the constraint $\Theta(\alpha\YP-\yijk)$ and integration limits
\beqn
0 < & \yijk & < \YP =\frac{1-\mu_Q}{1+\mu_Q} \nn \\
(1-\vijk)/2 < & z & < (1+\vijk)/2
\eeqn
where 
\beq
\vijk  = 
\frac{\sqrt{[2\mu_k^2+(1-\mu_k^2)(1-\yijk)]^2-4\mu_k^2}}
{(1-\mu_k^2)(1-\yijk)}.
\eeq 

For this case the subtraction term is 
\beqn
\langle s|\bV_{g q,Q}|s'\rangle && =
8\pi\mu^{2\ep}\alpha_S\CF \left\{
\frac{2}{1-\zj(1-\yijk)}
-\frac{\tvijk}{\vijk}\left[1+\zj+\ep(1-\zj) \eta \right]
\right\} \delta_{ss'}
\nn\\[.5em]
\eeqn
The correction pieces for the $\alpha$-dependent 
restricted range of the integral over $y$ can be performed with the help
of the transformation 
\beq
x = \YP-y+\sqrt{\Big(\YP-y\Big)\Big(\frac{1}{\YP}-y\Big)}
\eeq
as suggested in Appendix C.1 of ref.~\cite{Dittmaier:2000mb}. The result 
for the eikonal integral is,
\beqn
&&I^{\eik}(0,\mu_Q;\ep) =
\frac{1}{2 \ep^2}
-\frac{\ln(1-\mu_Q^2)}{\ep}\nn \\
&+&\Li_2(1-\mu_Q^2)
-\frac{5}{12} \pi^2 +\ln^2(1-\mu_Q^2)
\nn \\
      &+& \frac{1}{2} 
 \ln^2\Big(\frac{1-\YP^2+2 \xp \YP}{(1+\YP-\xp)(1-\YP+\xp)}\Big)
      -\ln^2\Big(\frac{1+\YP-\xp}{1+\YP}\Big) \nn \\
      &+&2 \Bigg[ \; \ln\Big(\frac{1+\YP}{2}\Big) \ln\Big(\frac{1-\YP+\xp}{1-\YP}\Big)
      +\; \ln\Big(\frac{1+\YP}{2 \YP} \Big) 
      \ln\Big(\frac{1-\YP^2+2 \xp \YP}{1-\YP^2} \Big) 
\nn \\
  &+&  \; \Li_{2}\Big(\frac{1-\YP}{1+\YP}\Big) 
    - \;  \Li_{2}\Big(\frac{1-\YP^2+2 \xp \YP}{(1+\YP)^2}\Big) \nn \\
  &+&  \; \Li_{2}\Big(\frac{1-\YP+\xp}{2}\Big)
   -\;   \Li_{2}\Big(\frac{1-\YP}{2}\Big)\Bigg]+\Oe{}\:
\eeqn
where
\beq
\xp= \YP(1-\alpha)+\sqrt{(1-\alpha)(1-\alpha\YP^2)}
\eeq
Setting $\alpha=1$ yields agreement with the appropriate limit of CDST, Eq.~(A.1).

The contribution from the collinear piece which has no 
soft singularity is  
\beqn
I^{\coll}(0,\mu_Q;\ep) &=&
\frac{3}{2 \ep} -3 \ln(1-\mu_Q)+\frac{9 + \eta}{2} 
-\frac{\mu_Q}{1-\mu_Q}
-2 \frac{\mu_Q (1-2 \mu_Q)}{(1-\mu_Q^2)} \nn \\
&-&\frac{3}{2}\Big[\ln(\alpha)+\YP(1-\alpha)\Big]+\Oe{}\:
\eeqn
After setting $\alpha=1$ and $\eta=1$,
this can be compared to CDST, Eq.~(5.35).
The total contribution for the integrated dipole is defined in 
Eq.~(\ref{eq:eikdiv}).

\end{document}